\documentclass[epjST]{svjour}
\usepackage{amsmath,amssymb,graphicx,bm, physics}
\usepackage{color}
\usepackage{ulem}
\usepackage{graphicx}
\usepackage{dcolumn}
\newcommand{\vek}[1]{\bm{\mathrm{#1}}}

\newcommand{\fmi}{\ifmmode\;\text{fm}^{-1}\else~fm$^{-1}$\fi}
\newcommand{\fm}{\;\text{fm}}
\newcommand{\MeV}{\;\text{MeV}}
\newcommand{\BCS}{\text{BCS}}

\DeclareMathOperator{\re}{Re}
\newcommand{\Vtilde}{\tilde{V}}

\newcommand{\kf}{k_{\text{F}}}

\newcommand{\kv}{\vek{k}}

\newcommand{\qv}{\vek{q}}

\newcommand{\pvec}{\vek{p}}
\newcommand{\vekk}{\vek{k}}

\newcommand{\vlowk}{\ensuremath{V_{\text{low}\,k}}}
\newcommand{\vsrg}{\ensuremath{V_{\text{srg}}}}

\newcommand{\Sec}[1]{Sec.~\ref{#1}}

\newcommand{\Eq}[1]{Eq.~\eqref{#1}}

\newcommand{\Refe}[1]{Ref.~\cite{#1}}
\newcommand{\Refs}[1]{Refs.~\cite{#1}}
\newcommand{\Fig}[1]{Fig.~\ref{#1}}
\newcommand{\Figs}[1]{Figs.~\ref{#1}}

\newcommand{\Gtemp}{\mathcal{G}}

\newcommand{\ncorr}{n_{\text{corr}}}
\newcommand{\nfree}{n_{\text{free}}}
\begin{document}

%%%%%%%%%%%%%%%%%%%%%%%%%%%%%%%%%%%%%%%%%%%%%%%%%%%%%%%%%%%%%%%%%%%%%%%%%%%%%%%
\title{Pairing in pure neutron matter} \author{S. Ramanan \inst{1}
  \and M. Urban \inst{2}}
\institute{
  Department of Physics, Indian
  Institute of Technology Madras, Chennai - 600036, India,\\
  \email{suna@physics.iitm.ac.in}
  \and
  Universit\'e Paris-Saclay, CNRS-IN2P3, IJCLab, 91405 Orsay cedex, France,\\
  \email{michael.urban@ijclab.in2p3.fr}
  }
  
%%%%%%%%%%%%%%%%%%%%%%%%%%%%%%%%%%%%%%%%%%%%%%%%%%%%%%%%%%%%%%%%%%%%%%%%%%%%%%%
\begin{abstract} 
{ We review the long standing problem of superfluid pairing in pure
  neutron matter. For the $s$-wave pairing, we summarize the state of the art
  of many-body approaches including different $nn$ interactions,
  medium polarization, short-range correlations and BCS-BEC crossover
  effects, and compare them with quantum Monte Carlo results at
  low-densities. We also address pairing in the $p$-wave, which
  appears at higher densities and hence has large uncertainties due to
  the poorly constrained interactions, medium effects and many-body
  forces.}
 \end{abstract}

\date{05/10/2020} \maketitle
%%%%%%%%%%%%%%%%%%%%%%%%%%%%%%%%%%%%%%%%%%%%%%%%%%%%%%%%%%%%%%%%%%%%%%%%%%%%%%%
\section{Introduction}
\label{sect:intro}
%%%%%%%%%%%%%%%%%%%%%%%%%%%%%%%%%%%%%%%%%%%%%%%%%%%%%%%%%%%%%%%%%%%%%%%%%%%%%%%
Superfluiditity in nuclei is nearly a 60 year old problem. However, a
satisfactory microscopic description of the phenomenon continues to
remain a challenge as the problem is marred by uncertainties in the
input interactions, both at the few body level and the medium
corrections. The possibility of neutron superfluidity was already
pointed out around 1960~\cite{Migdal1959,Ginzburg1964}. The
observational confirmation began with the discovery of
pulsars~\cite{Bell1968}, their connection to rotating neutron
stars~\cite{Gold1968} and the subsequent observation of glitches in
the period of rotation of these pulsars. Rotating neutron stars are
almost perfect clocks with a period of rotation that increases very
slowly with time. However, sometimes the period of rotation suddenly
decreases, followed by long relaxation times (over years) before it
returns to its pre-glitch value. Such glitches can be explained if one
allows for the existence of a superfluid phase in the inner crust of
the star through the mechanism of vortex
unpinning~\cite{Baym1969,Pines1985} (maybe one needs also
superfluidity in the core \cite{Andersson2012}). Further, the
existence of a superfluid state is crucial to explain the
observational data on
cooling~\cite{Yakovlev2004,Page2009,Anderson1975}.

The two-body interaction between two neutrons has attractive
components and while it is not sufficient to produce a bound
di-neutron state in free space, in the presence of other neutrons this
attraction leads to Cooper instability leading to the existence of a
superfluid phase with $s$-wave pairing, which typically exists in the
inner crust of neutron stars. The $NN$ interaction is attractive in
the spin triplet state as well that leads to $p$-wave pairing, and
such a phase is assumed to exist at higher densities in the outer
layers of the core of the star.

In addition to the physics of neutron star crusts, pairing plays a
crucial role in finite nuclei as well by contributing to extra
binding, for example the extra binding leading to an energy gap in
even-even nuclei compared to the quasi-particle spectrum of odd-$A$
nuclei or the even-odd staggering in binding
energy~\cite{Bohr1958,BohrMottelson1}. Close to the drip lines, large
even-odd-staggering has been observed in isotopes of C, Ne and
Mg~\cite{Fang2004,Ozawa2001,Hagino2011}.

In the literature, several extensive reviews already exist on the
subject of neutron star physics and superfluidity in both finite and
infinite systems~\cite{Chamel2008,Dean2003,Haskell2018,Sedrakian2019}. In the
present special topics issue, we aim to give a short overview of the
status of $s$-wave pairing, in particular screening and beyond-BCS
crossover effects, and of the outstanding questions of $p$-wave pairing.
%%%%%%%%%%%%%%%%%%%%%%%%%%%%%%%%%%%%%%%%%%%%%%%%%%%%%%%%%%%%%%%%%%%%%%%%%%%%%%%%
\section{Singlet pairing}
\label{sect:s-wave}
%%%%%%%%%%%%%%%%%%%%%%%%%%%%%%%%%%%%%%%%%%%%%%%%%%%%%%%%%%%%%%%%%%%%%%%%%%%%%%%%
\subsection{BCS gap equation}
\label{subsect:s-wave-bcs}
%%%%%%%%%%%%%%%%%%%%%%%%%%%%%%%%%%%%%%%%%%%%%%%%%%%%%%%%%%%%%%%%%%%%%%%%%%%%%%%%
In the case of an attractive interaction between fermions, the filled
Fermi sea becomes unstable with respect to the formation of Cooper
pairs. The starting point to study pairing is the BCS theory, where
the gap or the critical temperature is given by the BCS gap equation,
which in the $s$-wave spin-singlet ($^1S_0$) channel is given by~\cite{Schrieffer}:
\begin{equation}	
  \Delta(k) = -\frac{1}{\pi} \int_0^\infty dk^\prime \, k^{\prime \, 2}
  \, V(k, k') \frac{\Delta(k')
    \tanh\left(\frac{E(k')}{2T}\right)}{E(k')},
  \label{eq:BCS_eqn_sing}
\end{equation}
where $\Delta(k)$ is the momentum dependent gap, $V(k,k')$ is the
matrix element of the $s$-wave neutron-neutron ($nn$) interaction,
$E(k) = \sqrt{\xi^2(k) + \Delta^2(k)}$ is the quasi-particle energy
with $\xi(k) = \varepsilon(k)-\mu$ and $\varepsilon(k) = k^2/(2 m^*)$,
$m^*$ is the neutron effective mass, $T$ the temperature and $\mu$ the
chemical potential (including the mean-field energy shift). The
critical temperature $T_c$ is the highest temperature at which there
is a non-trivial solution for Eq.~\eqref{eq:BCS_eqn_sing}. At $T=T_c$,
the gap in $E(k')$ can be neglected and as a result
Eq.~\eqref{eq:BCS_eqn_sing} becomes a linear eigenvalue equation. In
the weak-coupling limit where $\Delta(\kf) \ll \mu$, the gap at zero
temperature is related to the BCS transition temperature by $T_c =
0.57 \, \Delta_{T = 0}(\kf)$. In the case of neutron matter, this
formula is a good approximation at all values of $\mu$, because the Fermi 
surface remains rather well defined. To simplify
the notation, we will from now on write $\Delta = \Delta(\kf)$.

In our
calculations we mostly use the renormalization group (RG) based interactions, 
$\vlowk$~\cite{Bogner2007} and $\vsrg$~\cite{Bogner2010}. They have an inherent scale ($\Lambda$ for $\vlowk$ and $\lambda$ for $\vsrg$) that sets the scale of decoupling between the
low and high momenta. Such a scale is arbitrary and observables should
be independent of this scale.
Within the simplest BCS approximation, i.e., employing the free-space
$nn$ interaction $V^0(k, k')$ and the free neutron mass $m^* =
m$, any realistic $nn$ interaction that reproduces the two-body
neutron phase shifts yields the same BCS gap \cite{Hebeler2007}. 

\begin{figure}
  \begin{center}
    \includegraphics[scale=0.72]{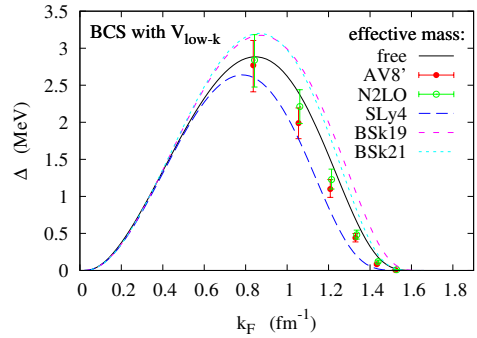}
    \end{center}
  \caption{BCS pairing gaps $\Delta = \Delta(\kf)$ obtained from
    \Eq{eq:BCS_eqn_sing} using the $\vlowk$ interaction (with a cutoff
    of $\Lambda = 2\fmi$) \cite{Bogner2007,Bogner2010}, as functions
    of $\kf$. The differences between the results are only due to the
    different effective masses $m^*$ used in the calculations: free
    mass ($m^* = m$, black solid line), effective masses from recent
    auxiliary-field diffusion Monte-Carlo calculations
    \cite{Buraczynski2019} using the AV8$'$+UIX interaction (red
    filled circles) and the chiral N2LO interaction (green empty
    circles), and effective masses from three different Skyrme
    parametrizations: SLy4 \cite{Chabanat1998} (blue long dashes),
    BSk19 and BSk21 \cite{Goriely2010} (purple short dashes and
    turquoise dots, respectively).}
  \label{fig:BCSgap}
\end{figure}

However, uncertainties arise already at the BCS level as soon as the
effective mass $m^*\ne m$ is included, since this affects the density
of states $N_0 = m^* \kf/\pi^2$, where $\kf = (3\pi^2 n)^{1/3}$ is the
Fermi momentum with $n$ the number density. Recent quantum Monte Carlo
(QMC) calculations \cite{Buraczynski2019} found that the neutron
effective mass drops only moderately with increasing density,
similar to what one gets with effective Gogny forces
\cite{Decharge1980,Chappert2008}, while effective Skyrme interactions
of the Saclay-Lyon family \cite{Chabanat1998} predict a stronger drop,
in contrast to those of the Bruxelles-Montreal family
\cite{Chamel2009,Goriely2010} which predict a slightly increasing effective
mass. In particular at higher densities beyond $\kf \approx 0.8\;
\fmi$ (corresponding to number densities above $0.017\fm^{-3}$ or mass
densities above $2.9\cdot10^{13}\;\text{g}/\text{cm}^3$) where the BCS
gap is maximum, the gap depends very sensitively on the density of
states, and therefore the different effective masses lead to
sizable uncertainties as can be seen in \Fig{fig:BCSgap}.
%%%%%%%%%%%%%%%%%%%%%%%%%%%%%%%%%%%%%%%%%%%%%%%%%%%%%%%%%%%%%%%%%%%%%%%%%%%%%%%%
%%%%%%%%%%%%%%%%%%%%%%%%%%%%%%%%%%%%%%%%%%%%%%%%%%%%%%%%%%%%%%%%%%%%%%%%%%%%%%%

At lower densities, the effective mass is close to the free one, and
the gap is less sensitive to it. In this region, the uncertainties
come mainly from corrections beyond the BCS approximation. These will
be addressed in the following subsections.
%%%%%%%%%%%%%%%%%%%%%%%%%%%%%%%%%%%%%%%%%%%%%%%%%%%%%%%%%%%%%%%%%%%%%%%%%%%%%%%
\subsection{Screening corrections}
\label{subsect:screening}
\begin{figure}
  \begin{center}
    \includegraphics[scale=0.57, clip = true]{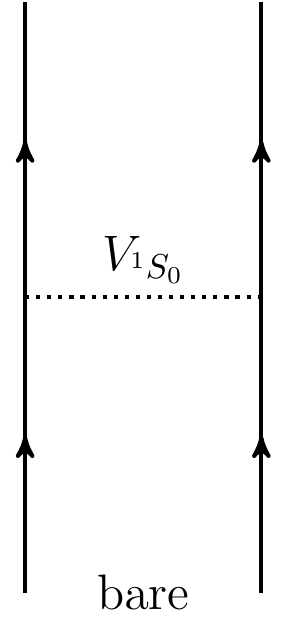}
    \hspace*{0.2in}
    \includegraphics[scale=0.57, clip = true]{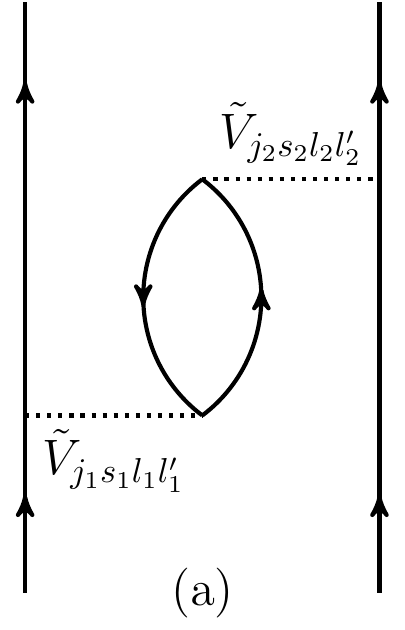}
    \hspace*{0.2in}
    \includegraphics[scale=0.57, clip = true]{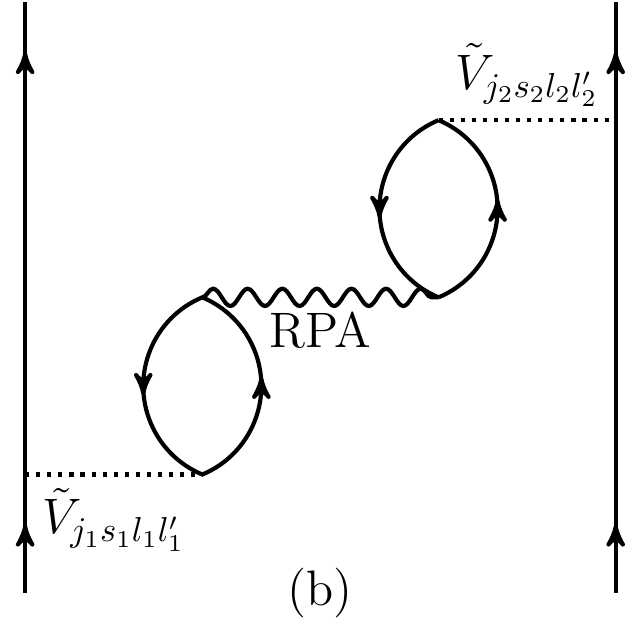}
    \hspace*{0.2in}
    \includegraphics[scale=0.57, clip = true]{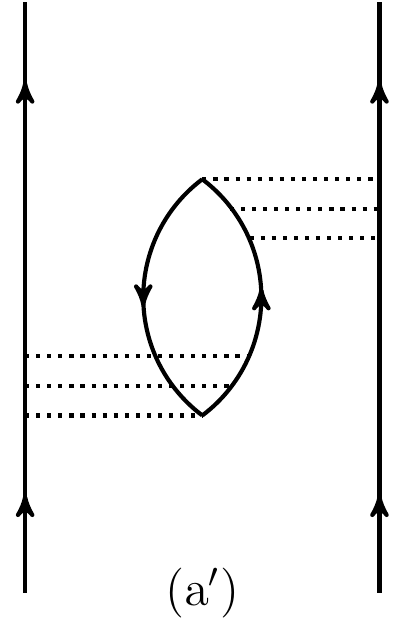}
  \end{center}
  \caption{In the medium, the bare pairing interaction (leftmost
    diagram) is modified by the screening corrections (a) and
    (b). Diagram (a$'$) illustrates the resummation of ladders in the
    3p1h vertices of diagram (a) implicitly assumed in the derivation
    of the GMB result \cite{Gorkov1961}.}
  \label{fig:vind}
\end{figure}

%%%%%%%%%%%%%%%%%%%%%%%%%%%%%%%%%%%%%%%%%%%%%%%%%%%%%%%%%%%%%%%%%%%%%%%%%%%%%%%
It is well known that corrections beyond the BCS approximation due to
density and spin-density fluctuations that the neutrons create in the
surounding medium are very important. Such corrections are called
medium polarization or screening effects, since they are analogous to
the screening of the Coulomb interaction. They can be taken into
account in the gap equation by adding the induced interaction to the
bare $nn$ interaction, so that,
\begin{equation}
  V(k,k^\prime) = V^0(k,k^\prime)+V^{(a)}(k,k^\prime)+V^{(b)}(k,k^\prime)\,,
  \label{eq:V0ab}
\end{equation}
where the induced interactions, $V^{(a)}$ and $V^{(b)}$, are as seen
in \Fig{fig:vind}. In this figure, diagram (a) allows for one
particle-hole (ph) bubble insertion while diagram (b) sums the ph
bubble series (random-phase approximation, RPA, represented by wavy
lines).
%%%%%%%%%%%%%%%%%%%%%%%%%%%%%%%%%%%%%%%%%%%%%%%%%%%%%%%%%%%%%%%%%%%%%%%%%%%%%%%
%%%%%%%%%%%%%%%%%%%%%%%%%%%%%%%%%%%%%%%%%%%%%%%%%%%%%%%%%%%%%%%%%%%%%%%%%%%%%%%
In these diagrams, the interaction $\Vtilde$ shown by the dotted lines
is meant to be antisymmetrized, $\langle 12|\Vtilde|34\rangle =
\langle 12|V|34\rangle-\langle 12|V|43\rangle$, i.e., it includes also
the exchange graphs which are not drawn.

\begin{figure}
  \begin{center}
    \includegraphics[scale=0.72]{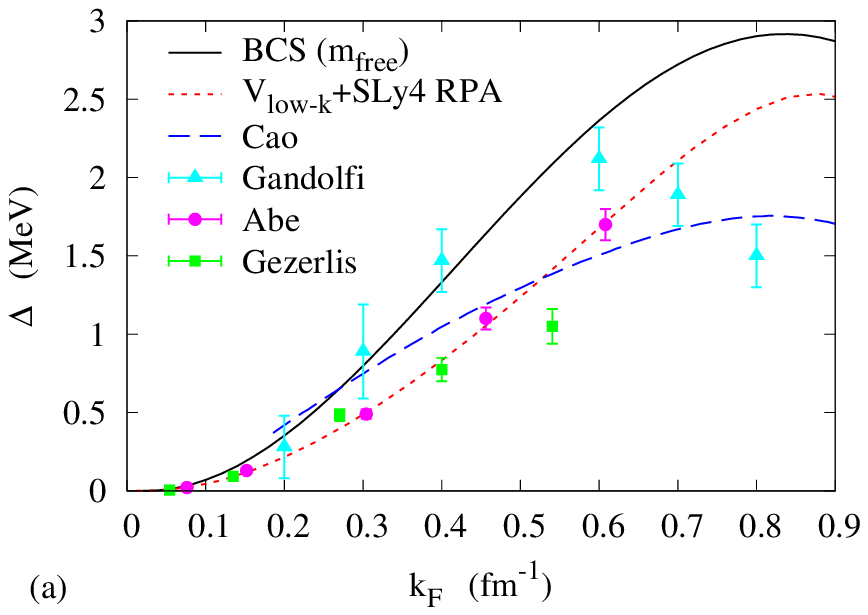}
    \includegraphics[scale=0.72]{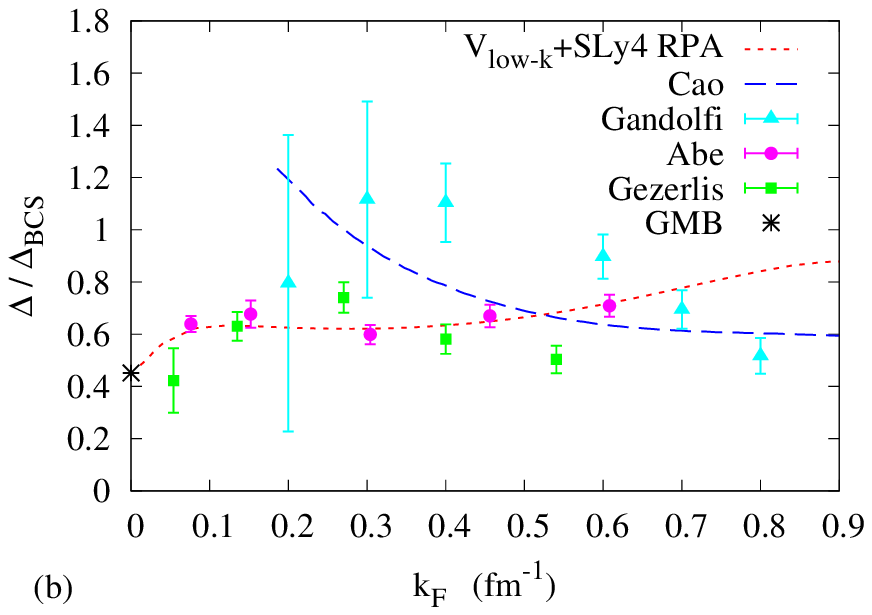}
  \end{center}
  \caption{(a) Screening and QMC results for the gap in neutron matter
    as a function of the Fermi momentum $\kf$. The blue dashes and red
    dots are the final results of the screening calculations of Cao et
    al. \cite{Cao2006} and of our own work \cite{Urban2020},
    respectively. The turquoise triangles, purple points, and green
    squares are QMC results of Gandolfi et al. \cite{Gandolfi2008},
    Abe and Seki \cite{Abe2009}, and Gezerlis and Carlson
    \cite{Gezerlis2010}, respectively. For comparison, the BCS result
    $\Delta_{\BCS}$ obtained without effective mass is shown as the
    black line (same as in \Fig{fig:BCSgap}). (b) Same data as in (a)
    but normalized to $\Delta_{\BCS}$. The GMB result
    \cite{Gorkov1961} is shown as the black star.}
  \label{fig:screen-QMC}
\end{figure}

There have been many attempts to calculate the induced interactions in
the
literature~\cite{Wambach1993,Schulze1996,Shen2003,Shen2005,Cao2006,Ramanan2018,Urban2020}. Especially
the earlier calculations \cite{Wambach1993,Schulze1996} found an
extremely strong suppression of the gap. However, since the work by
Cao et al.~\cite{Cao2006} a consensus seems to emerge that the gap is
not too strongly reduced. This is shown in \Fig{fig:screen-QMC} which
summarizes more recent screening and QMC results.
%%%%%%%%%%%%%%%%%%%%%%%%%%%%%%%%%%%%%%%%%%%%%%%%%%%%%%%%%%%%%%%%%%%%%%%%%%%%%%%
%%%%%%%%%%%%%%%%%%%%%%%%%%%%%%%%%%%%%%%%%%%%%%%%%%%%%%%%%%%%%%%%%%%%%%%%%%%%%%%
In \Fig{fig:screen-QMC}(b) we also show the result
$\Delta/\Delta_{\BCS}=(4e)^{-1/3}\approx 0.45$ (black star) obtained
long ago by Gor'kov and Melik-Barkhudarov (GMB) \cite{Gorkov1961},
which should become valid in the limit $|\kf a_{nn}| \ll 1$, with
$a_{nn} \approx -18.5\fm$ the $nn$ scattering length.

It is seen that the two screening calculations
\cite{Cao2006,Urban2020} do still not quite agree with each other. We
will come back to a more detailed discussion of these calculations
below. The QMC calculations, which are supposed to be, up to numerical
limitations, exact solutions of the many-body problem, show only a
moderate suppression. The gaps of \Refe{Gandolfi2008} (turquoise
triangles), were obtained with the auxiliary-field diffusion Monte
Carlo technique using the Argonne V8$'$ $nn$ interaction (AV8$'$) and the
Urbana IX three-body force (UIX) and is not significantly reduced
compared to $\Delta_{\BCS}$ up to $\kf\sim 0.6\fmi$, but the error
bars are huge. The green points of \Refe{Abe2009} were obtained within
a method based on the discretization of the Hamiltonian on a lattice
(determinantal quantum Monte Carlo). The interaction used in this
calculation is much simpler, as it includes only the leading and
next-to-leading orders (NLO) of pionless effective field theory (EFT),
and is only valid at low momenta, i.e., low densities. These gaps are
reduced by an almost constant factor of about $0.6-0.7$ compared to
$\Delta_{\BCS}$ (see \Fig{fig:screen-QMC}(b)). The almost perfect
agreement of these results with the red dashed curve is probably
accidental. A similar behavior was found in \Refe{Gezerlis2010} using
the AV4 interaction within the variational and subsequent Green's
function Monte-Carlo method. At very low densities, these results tend
(within the error bars) towards the GMB limit. According to
\Refe{Gezerlis2010}, the discrepancy between \Refs{Gandolfi2008} and
\cite{Gezerlis2010} might be due to the less optimized wave function
used in \Refe{Gandolfi2008}.

\begin{figure}
  \begin{center}
    \includegraphics[scale=0.72]{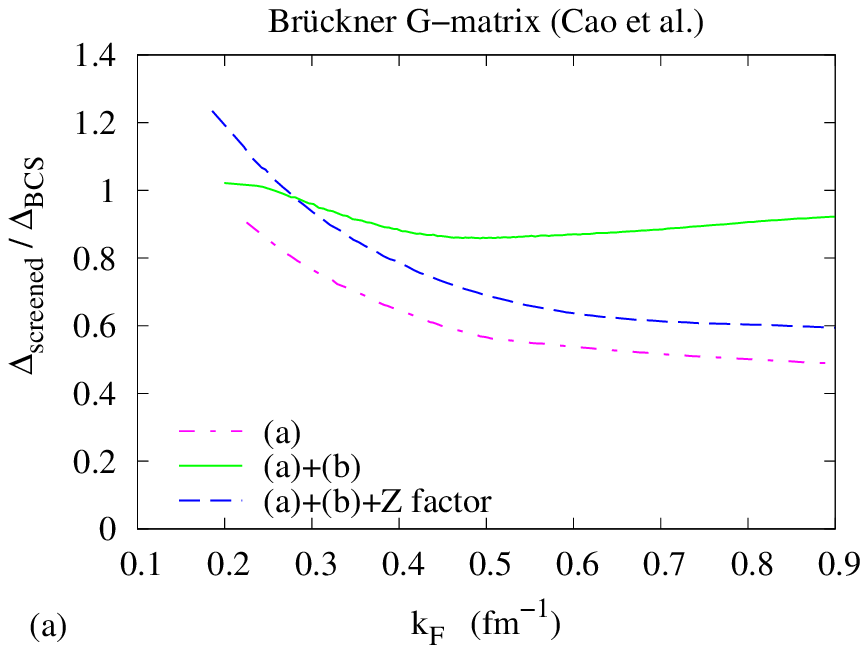}
    \includegraphics[scale=0.72]{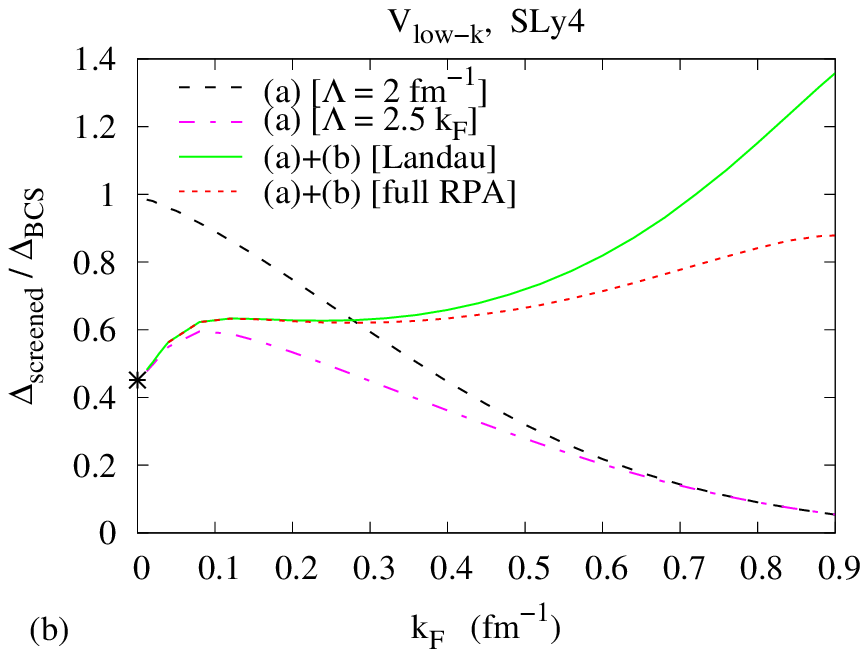}
  \end{center}
  \caption{Results for the screened gap in neutron matter as a
    function of the Fermi momentum $\kf$ at different steps of the
    screening calculations of \Refe{Cao2006} (a) and of our own
    calculations \cite{Ramanan2018,Urban2020} (b). See text for
    details.}
  \label{fig:screen-low}
\end{figure}

Let us now discuss in some more detail the screening calculations. In
\Fig{fig:screen-low}
%%%%%%%%%%%%%%%%%%%%%%%%%%%%%%%%%%%%%%%%%%%%%%%%%%%%%%%%%%%%%%%%%%%%%%%%%%%%%%%
%%%%%%%%%%%%%%%%%%%%%%%%%%%%%%%%%%%%%%%%%%%%%%%%%%%%%%%%%%%%%%%%%%%%%%%%%%%%%%%
we display again the ratios of screened gaps to our reference curve
$\Delta_{\BCS}$ which is the BCS gap with the free neutron mass (black
solid line in \Fig{fig:BCSgap}), including the results obtained at
intermediate steps on the way to the final results. Figure
\ref{fig:screen-low}(a) summarizes the neutron-matter results of
\Refe{Cao2006}. In that work, the 3p1h vertices $\tilde{V}$ (dotted
lines in \Fig{fig:vind}) are the Br\"uckner G matrix. Up to the
projection on the $^1S_0$ wave, diagram (a) can be schematically
written as
\begin{equation}
  V^{(a)} = \frac{\pi}{2}\sum_{\pvec\sigma}\Vtilde\,
  \frac{n(\pvec-\frac{\qv}{2})-n(\pvec+\frac{\qv}{2})}
       {\varepsilon(\pvec+\frac{\qv}{2})-\varepsilon(\pvec-\frac{\qv}{2})}\,
       \Vtilde\,,
\end{equation}
where we have omitted all momentum and spin labels of $\Vtilde$. Here,
$\pvec$ and $\sigma$ are the momentum and spin labels that are summed
over in the ph loop and $\qv = \kv-\kv'$ is the momentum transfer. The
occupation numbers can be safely approximated by step functions
$n(\pvec)=\theta(\kf-|\pvec|)$. Also, as it is usually done, the
static approximation is made, i.e., the energy transfer in the ph
bubble is neglected. To simplify this complicated expression, the
authors of \Refe{Cao2006} replaced $\Vtilde$ by its average value
$\langle\Vtilde\rangle$, where the averaging is done around the Fermi surface, so that it could be taken out of the sum,
which then gives
\begin{equation}
  V^{(a)} = -\frac{\pi}{2} \langle\Vtilde\rangle^2\,\Pi^0(q/\kf)\,,
  \label{Va-schematic}
\end{equation}
where $\Pi^0(\tilde{q})$ is the static Lindhard function
[$\Pi^0(0,\tilde{q})$ in Eq.~(12.46b) of \cite{FetterWalecka}, with
  $m$ replaced by $m^*$] with $\tilde{q} = q/\kf$. The subsequent
projecting of $V^{(a)}$ on the $s$ wave finally amounts to averaging
the Lindhard function over the angle between $\kv$ and $\kv'$, i.e.,
over $q$ in the range $|k-k'| \le q \le k+k'$. Since $\Pi^0 < 0$, the
induced interaction $V^{(a)}$ is repulsive and therefore reduces
(screens) the bare interaction $V^{0}$. The solution of the gap
equation with $V^0+V^{(a)}$ is shown as the purple dash-dot line in
\Fig{fig:screen-low}(a). We see that the screening disappears at low
densities, which is easily understood since $\Pi^0 \propto k_F$ in
\Eq{Va-schematic}.

The next step is to include also diagram (b) of \Fig{fig:vind}. Using
the Landau approximation, the residual ph interaction is approximated
as ${\cal V} = f_0 + g_0 \, \bm{\sigma}_1 \cdot \bm{\sigma}_2$, with
$f_0$ and $g_0$ the Landau parameters, and thereby the RPA series can
be separately summed in the $S = 0$ and $S = 1$ channels, where $S$
denotes the total spin of the ph excitation. Then, the inclusion of
diagram (b) modifies \Eq{Va-schematic} to
\begin{equation}
  V^{(a)}+V^{(b)} = -\frac{\pi}{2} \langle\Vtilde\rangle^2
  \Big(\frac{3}{2}\Pi_{S=1}-\frac{1}{2}\Pi_{S=0}\Big)\,,
  \label{Vab-schematic}
\end{equation}
with
\begin{equation}
  \Pi_{S=0} = \frac{\Pi^0}{1-f_0\Pi^0}\,,\qquad
  \Pi_{S=1} = \frac{\Pi^0}{1-g_0\Pi^0}\,.
\end{equation}
Notice that, in dilute neutron matter, $g_0 > 0$ and $f_0 <
0$. Together with $\Pi^0 < 0$, this implies that, with increasing
density, the RPA enhances the attractive $S=0$ contribution in
\Eq{Vab-schematic} while it reduces the repulsive $S=1$
contribution. The net effect of diagram (b) is therefore that the gap
(green solid line in \Fig{fig:screen-low}(a)) is much less screened
than with diagram (a) only.

To obtain the final result of \Refe{Cao2006}, another effect was taken
into account. Namely, the energy dependence of the self-energy
$\Sigma(k,\omega)$ computed in Br\"uckner theory (Fig.~1(b)
of~\cite{Baldo2000}) leads to a reduction of the quasiparticle weight
$Z(k) = 1/(1-\partial\Sigma/\partial\omega)$. This effect can be
accounted for by introducing a factor of $Z(k)Z(k')$ on the right-hand
side of the gap equation (\ref{eq:BCS_eqn_sing}), which then yields
the final result shown in \Fig{fig:screen-QMC} and in
\Fig{fig:screen-low}(a) as the blue dashed lines.

In spite of the reasonable agreement with the QMC results at high density,
the increase of the gap at low-densities ($\kf \lesssim 0.27 \, \fmi$) looks 
somewhat suspicious. Furthermore, besides the approximation of $\Vtilde$ by its
average $\langle\Vtilde\rangle$ mentioned above, \Refe{Cao2006} used
the Babu-Brown theory \cite{Babu1973} to determine the Landau
parameters in a self-consistent way, with the aim to avoid the
liquid-gas instability in low-density symmetric nuclear
matter. However, the validity of this argument may be questioned as
the liquid-gas instability exists.

For these reasons, the screening problem was reconsidered by the
authors in \Refs{Ramanan2018} and \cite{Urban2020}. As input $nn$
interaction, $V^0(k, k')$ in Eq.~\eqref{eq:V0ab}, as well as in the
antisymmetrized 3p1h vertices, we use $\vlowk$. For the ph interaction in the RPA, as
well as in the calculation of the effective mass $m^*$, we use for
simplicity a phenomenological Skyrme energy-density functional (SLy4
in the present example). No further approximations are made, and in
particular, the full momentum dependence of the 3p1h vertices is taken
into account when summing over the loop momenta.

Starting with diagram (a) in \Fig{fig:vind}, computed with the
$\vlowk$ interaction obtained for a common choice of the cutoff
$\Lambda = 2\fmi$, one obtains the gap shown in
\Fig{fig:screen-low}(b) as the black dashed line. As already observed
in~\cite{Shen2005,Cao2006}, the screening vanishes and the BCS result
is recovered in the limit $\kf\to 0$, in contradiction to the GMB
result. In fact, the GMB result \cite{Gorkov1961} is also based on
diagram (a) (since all other diagrams can be neglected in the limit
$\kf a\to 0$), but with a subtle difference: In the 3p1h vertices,
one has to use the scattering length (i.e., the full T matrix), and
not just the bare interaction $V$ used in diagram (a). This amounts to
implicitly summing ladders to all orders in the 3p1h vertices, as
shown in diagram (a$'$).

Making use of the RG flow of the $\vlowk$ interaction, a simple way to
solve this problem was suggested in \Refe{Ramanan2018}. First, notice
that, when decreasing the cutoff $\Lambda$, the RG flow guarantees
that the scattering length $a_{nn}$ remains constant by increasing the
matrix elements of the interaction as $V \approx
(m/a-2m\Lambda/\pi)^{-1}$. In this way, the interaction becomes more
and more perturbative in the sense that the Born term is already a
good approximation to the full T matrix. Second, the RG evolution of
$\vlowk$ leaves the BCS gap independent of the cutoff $\Lambda$, as
long as $\Lambda\gtrsim 2.5 \kf$. So, it is preferable to scale the
cutoff with $\kf$, using at each density the lowest permissible cutoff
$\Lambda = 2.5 \kf$. Calculating diagram (a) of \Fig{fig:vind} with 
this prescription, one obtains the result shown in \Fig{fig:screen-low}(b)
as the purple dash-dotted line, which indeed reproduces the GMB result
(black star) in the limit $\kf\to 0$.

At higher densities, the RPA corrections [\Fig{fig:vind}(b)] become
important. In the case that the ph interaction is of the Skyrme type,
it is rather straightforward to resum the RPA bubble series
exactly~\cite{Urban2020,Garcia1992,Pastore2015}. This gives our final
result shown as the red dotted lines in \Figs{fig:screen-QMC} and
\ref{fig:screen-low}(b). As discussed above, the inclusion of
\Fig{fig:vind}(b) strongly reduces the screening effect of diagram
(a).

If we use in diagram (b) instead of the full RPA the Landau
approximation, as it was done in
\Refs{Shen2005,Cao2006,Ramanan2018,Ding2016}, we obtain the green
solid line shown in \Fig{fig:screen-low}. Comparing this result with
the red dashed line, one concludes that the Landau approximation is
only valid for $\kf \lesssim 0.4\fmi$. Beyond this density, it
overestimates the effect of the RPA and, for $\kf>0.7\fmi$, it
even predicts anti-screening (i.e., the gap is enhanced) because of
the large values of the Landau parameters. Anti-screening was already
found long ago in \Refe{Schulze1996}, but only at much higher
densities ($\kf\gtrsim1.3\fmi$). The strong anti-screening effect
found in \cite{Ramanan2018} within the Landau approximation at higher
density is absent or strongly suppressed within the full RPA
calculation \cite{Urban2020}.

\begin{figure}
  \begin{center}
    \includegraphics[scale = 0.72]{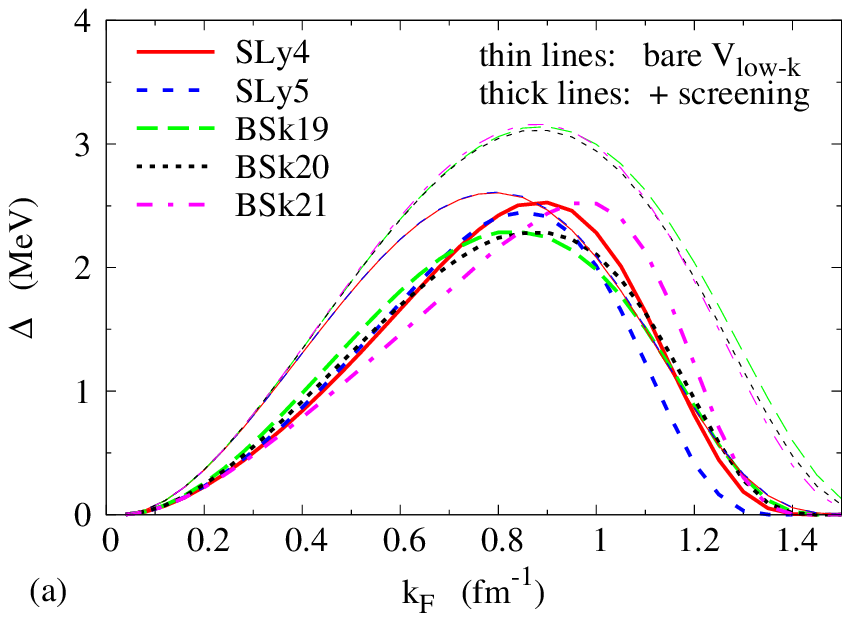}
    \includegraphics[scale = 0.72]{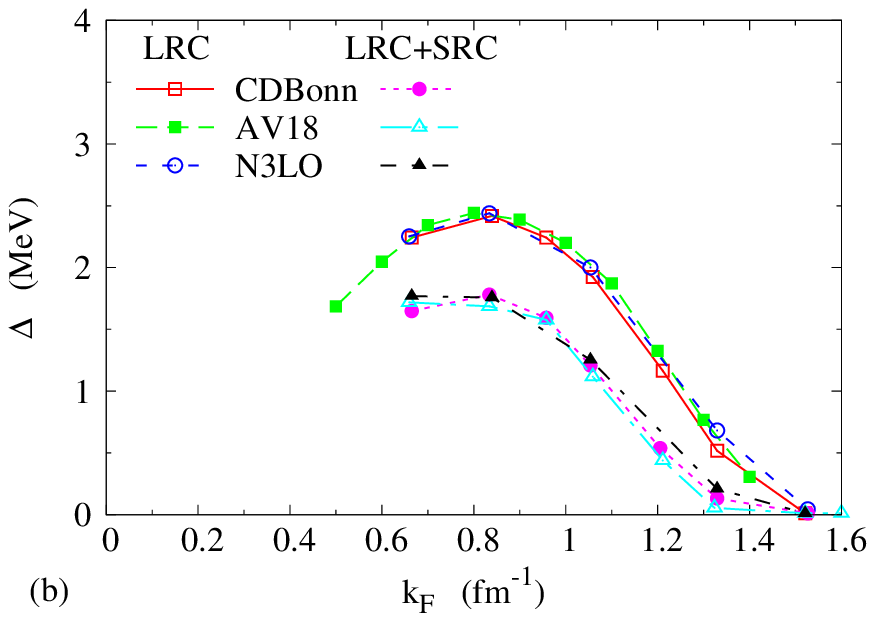}
   \end{center}
  \caption{Behaviour of the singlet gap $\Delta$ versus $\kf$. (a)
    Results of \Refe{Urban2020} with (thick lines) and without (thin
    lines) medium polarization corrections for different Skyrme
    parameterizations used in the effective mass and in the RPA bubble
    summation. (b) Results of \Refe{Ding2016} including effects of
    medium polarization (=long-range correlations, LRC) as well as
    short-range correlations (SRC).}
  \label{fig:screen-high}
\end{figure}

So far we have concentrated only on the low-density region with
$\kf<0.9\fmi$. At higher densities, as we have seen in
\Sec{subsect:s-wave-bcs}, the effective mass $m^*$ leads to large
uncertainties. Similarly, for the screening diagram (b), uncertainties
arise from the Landau parameters $f_0$ and $g_0$ and more generally,
if one goes beyond the Landau approximation, from the ph residual
interactions. Hence, in \cite{Urban2020}, we repeated the calculations
with a couple of different Skyrme parametrizations. All screened
results shown in \Fig{fig:screen-high}(a)
%%%%%%%%%%%%%%%%%%%%%%%%%%%%%%%%%%%%%%%%%%%%%%%%%%%%%%%%%%%%%%%%%%%%%%%%%%%%%%%
%%%%%%%%%%%%%%%%%%%%%%%%%%%%%%%%%%%%%%%%%%%%%%%%%%%%%%%%%%%%%%%%%%%%%%%%%%%%%%%
(thick lines) were computed with the full RPA and with the density
dependent cutoff $\Lambda=2.5 \,\kf$ for $\kf<0.8\fmi$, while we kept
$\Lambda=2\fmi$ constant for $\kf\geq 0.8\fmi$ since this
cutoff gives the correct BCS gap in the whole density range, and with
larger values of $\Lambda$ the advantage of the soft $\vlowk$ interactions would be
lost. Surprisingly, when screening is included, the dependence on the
choice of the Skyrme interaction is weaker than without screening. In
particular, for all the considered Skyrme forces the maximum of the
screened gap lies now between $2.3$ and $2.5\MeV$.

These results can be compared with a calculation based on the
self-consistent Green's function theory \cite{Ding2016}. Here, the
energy and momentum dependent single-particle self-energy
$\Sigma(k,\omega)$ is computed in ladder approximation, whereby all
propagators are themselves dressed ones. This approach accounts
automatically for the short-range correlations created by the
realistic (hard) $nn$ interactions, but not for screening, which
corresponds to long-range correlations. In \Refe{Ding2016},
screening was in fact only included in an approximate way, by adding
$V^{(a)}+V^{(b)}$ using the same approximations as in \Refe{Cao2006}
(see above). The results, obtained with three different bare $nn$
interactions (AV18, CDBonn, and the chiral N3LO interaction) are shown
in \Fig{fig:screen-high}(b). As long as only screening is
included (red, green and blue points), the maximum of the gap is again
about $2.5\MeV$, but the density where it tends to zero is clearly
higher than in our screening calculations
(\Fig{fig:screen-high}(a)). However, one should keep in mind that for
the momentum transfers needed in this density region neither the
Landau approximation nor the full Skyrme ph interaction can be
considered to be reliable.

The effect of short-range correlations is closely related to the $Z$
factors included in the gap equation in \Refe{Cao2006}. However, using the full
spectral functions as done in \cite{Ding2016} and not just the
quasiparticle peak, it becomes somewhat more sophisticated. Taking into
account the short-range correlations in addition to the screening 
(black, purple, and turquoise points in \Fig{fig:screen-high}(b)), the maximum 
gap is further
reduced to $\approx 1.8\MeV$. Another observation is that also
the density where the $^1S_0$ gap goes to zero is reduced. Apparently this effect is important and should be studied also at lower densities,
along with a more complete treatment of the screening. Short-range 
correlations can also be included via the correlated basis 
function method that once again leads to a suppression of the BCS gap~\cite{Pavlou2017},
but this technique will not be discussed in this short review. 
%%%%%%%%%%%%%%%%%%%%%%%%%%%%%%%%%%%%%%%%%%%%%%%%%%%%%%%%%%%%%%%%%%%%%%%%%%%%%%%%
\subsection{BCS-BEC Crossover}
\label{subsect:bcs-bec}
%%%%%%%%%%%%%%%%%%%%%%%%%%%%%%%%%%%%%%%%%%%%%%%%%%%%%%%%%%%%%%%%%%%%%%%%%%%%%%%%
The BCS-BEC crossover has attracted a lot of attention in the last two
decades, especially because its experimental realization in ultracold
trapped atoms. In these experiments, one can change the interatomic
interaction by varying the magnetic field, in such a way that the
system passes continuously from a BCS superfluid in the case of weakly
attractive interactions, through a resonance where the scattering
length $a$ diverges (unitary limit), to a Bose-Einstein condensate
(BEC) of bound dimers. For recent reviews emphasizing the analogies
between ultracold atoms and nuclear and neutron matter, see
\cite{CalvaneseStrinati2018,Ohashi2020}.

Of course, in nuclear systems, the interaction cannot be changed. In
this case, the crossover can be realized with changing density. Very
dilute symmetric nuclear matter will form a BEC of deuterons which,
with increasing density, goes continuously over into a BCS state with
$pn$ Cooper pairs \cite{Baldo1995}. In neutron matter, however, a
BCS-BEC crossover does not exist, because there is no bound dineutron
state. But the $nn$ scattering length is unusually large in the $s$
wave, signalling a nearly bound state. Hence, the Cooper pairs in
dilute neutron matter have a relatively small size (coherence length),
comparable to the average distance between particles
\cite{Matsuo2006,Margueron2007}.

\begin{figure}
  \begin{center}
    \includegraphics[scale = 0.72]{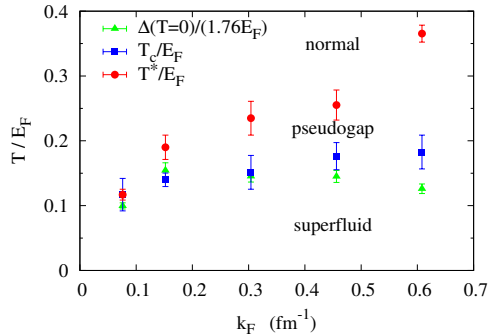}
   \end{center}
  \caption{QMC phase diagram of \Refe{Abe2009} displaying the critical
    temperature $T_c$ (blue squares) and the pair dissociation
    temperature $T^*$ (red points) in units of the Fermi energy
    $E_F=\kf^2/(2m)$ versus $\kf$. For comparison, the critical
    temperatures one would get from the BCS relation $T_c =
    0.57\Delta(T=0)$ are shown, too (green triangles).}
  \label{fig:TstarTc}
\end{figure}

In this case, similar to the situation when there is a true bound
state, the temperature $T^*$ where pairs dissociate can be higher than
the superfluid critical temperature $T_c$ where the pairs undergo
Bose-Einstein condensation. This can be seen in \Fig{fig:TstarTc}
%%%%%%%%%%%%%%%%%%%%%%%%%%%%%%%%%%%%%%%%%%%%%%%%%%%%%%%%%%%%%%%%%%%%%%%%%%%%%%%
%%%%%%%%%%%%%%%%%%%%%%%%%%%%%%%%%%%%%%%%%%%%%%%%%%%%%%%%%%%%%%%%%%%%%%%%%%%%%%%
which shows the QMC results of \Refe{Abe2009} for $T^*$ and $T_c$ as
functions of $\kf$. For a better visibility of the low-density
results, we have divided $T^*$ and $T_c$ by the Fermi energy $E_F =
\kf^2/(2m)$. The region between $T_c$ and $T^*$ is called the
pseudogap phase because, although there is no true gap, there exists a
suppression of the level density at $\omega = 0$ (energy measured
relative to the chemical potential $\mu$) because of the energy needed
to break a pair.

\begin{figure}
  \begin{center}
\includegraphics[angle = 0, width = 10cm, clip = true]{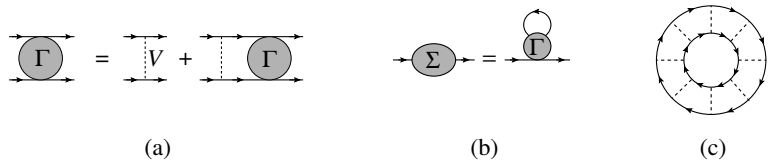}
\end{center}
\caption{Diagrams for the T matrix (a), the self energy (b), and the
  thermodynamic potential (c) in ladder approximation.}
\label{fig:feyn}
\end{figure}

In the pseudogap region, it is usually a bad approximation to compute
the density from the uncorrelated occupation numbers
\begin{equation}
\nfree = 2\int \frac{d^3k}{(2\pi)^3} f(\xi(\vekk))\,,
\label{eq:rho_free}
\end{equation}
where the factor of $2$ accounts for the spin degeneracy and $f(\xi) =
1/(e^{\xi/T}+1)$ is the Fermi function. Taking into account the
density corresponding to the correlated pairs is crucial to get the
correct result for $T_c$ in the BEC limit. This is done by the
Nozi\`eres-Schmitt-Rink (NSR) approach \cite{Nozieres1985}, which writes
\begin{equation} 
  n = \nfree + \ncorr\,.
\label{eq:NSR_decomp}
\end{equation}
The correlated density $\ncorr$ is calculated to first order in
the self-energy $\Sigma$ (in the imaginary time
formalism~\cite{FetterWalecka}),
\begin{equation}
  \ncorr = 2 \int\! \frac{d^3k}{(2\pi)^3} \frac{1}{\beta}
  \sum_{\omega_n} \big(\Gtemp_0(\vek{k}, \omega_n)\big)^2
[\Sigma(\vek{k}, i\omega_n) - \re \Sigma(\vek{k},\xi(\vek{k})]\,,
 \label{eq:rhocorr}
\end{equation}
where $\omega_n$ are the Matsubara frequencies and $\Gtemp_0$ is the
uncorrelated single-particle Green's function. The self-energy
$\Sigma$ is calculated within the ladder approximation as shown in
\Fig{fig:feyn}(a) and (b).
%%%%%%%%%%%%%%%%%%%%%%%%%%%%%%%%%%%%%%%%%%%%%%%%%%%%%%%%%%%%%%%%%%%%%%%%%%%%%%%
%%%%%%%%%%%%%%%%%%%%%%%%%%%%%%%%%%%%%%%%%%%%%%%%%%%%%%%%%%%%%%%%%%%%%%%%%%%%%%%
In the original NSR paper \cite{Nozieres1985}, the correlated density
is obtained as the derivative with respect to $\mu$ of the
thermodynamic potential represented by diagram (c) in \Fig{fig:feyn},
which is equivalent to keeping the self-energy only to first order in
\Eq{eq:rhocorr} \cite{CalvaneseStrinati2018}. However, the subtraction
of the on-shell self energy $\Sigma(\vek{k},\xi(\vek{k}))$ in
\Eq{eq:rhocorr} is absent in the original NSR approach. It is
necessary as this term is already taken into account in $\Gtemp_0$ via
the quasiparticle
energy~\cite{Zimmermann1985,Schmidt1990,Jin2010,Ramanan2013}.

The correlated density was calculated in~\cite{Ramanan2013} using the
$\vlowk$ interaction. In order to accommodate the non-local
interaction, the authors expressed the correlated density in the basis
that diagonalizes $V \bar{G}^{(2)}_0$, where $\bar{G}^{(2)}_0$ is the
two particle retarded Green's
function. In~\cite{Ramanan2018,Urban2020} the bare interaction was
augmented by the induced interaction as discussed in
\Sec{subsect:screening}. A similar calculation using a separable
interaction instead of $\vlowk$, but without screening corrections,
was done in \cite{Tajima2019}. In all these calculations, the
subtraction term was approximated by the first-order Hartree-Fock
(HF) self-energy. For a detailed comparison of different subtraction
prescriptions, see \cite{Durel2020}.

%%%%%%%%%%%%%%%%%%%%%%%%%%%%%%%%%%%%%%%%%%%%%%%%%%%%%%%%%%%%%%%%%%%%%%%%%%%%%%%
\begin{figure}
 \begin{center}
  \includegraphics[angle = 0, scale = 0.35, clip = true]{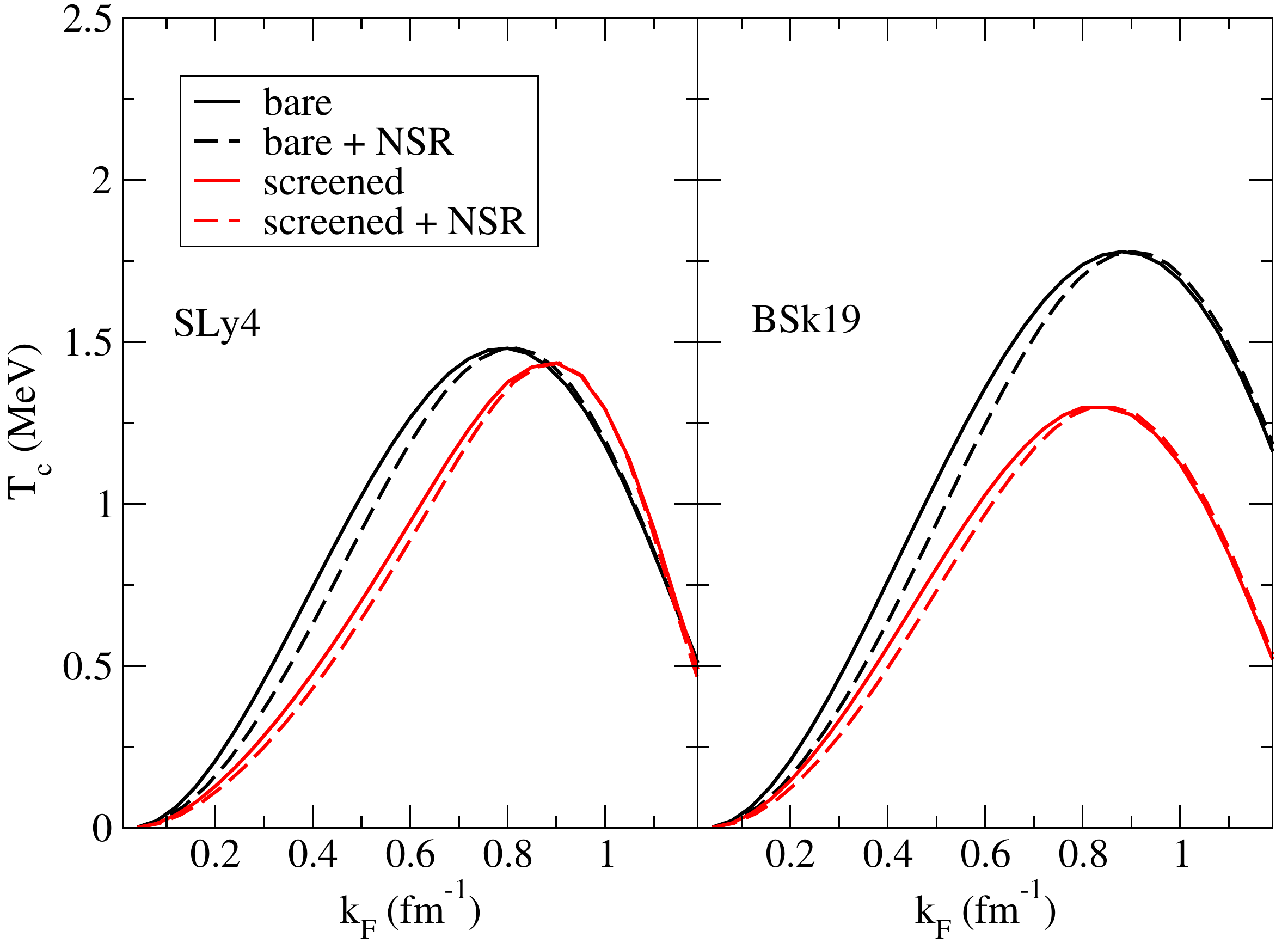}
 \end{center}
 \caption{Critical temperature $T_c$ with (red) and without (black)
   screening corrections, as a function of $k_F$ computed with
   (dashes) and without (solid lines) the NSR correction to the
   density, for two different Skyrme parmetrizations used in the
   calculation of the effective mass and of the screening corrections
   $V^{(a)}+V^{(b)}$ \cite{Urban2020}.}
	\label{fig:NSR-compare}
\end{figure}
%%%%%%%%%%%%%%%%%%%%%%%%%%%%%%%%%%%%%%%%%%%%%%%%%%%%%%%%%%%%%%%%%%%%%%%%%%%%%%%
Fig.~\ref{fig:NSR-compare} shows the effect of including the
correlated density on the density dependence of the transition
temperature \cite{Urban2020}. The black lines include the effective
mass $m^*$, computed with SLy4 (left panel) and BSk19 (right panel),
while the red lines include also the screening effects
$V^{(a)}+V^{(b)}$ (calculated with the same Skyrme force as
$m^*$). For the solid lines, $\kf$ was computed with $\nfree$, while
for the dashed lines, $\kf$ was computed with the NSR density
$\nfree+\ncorr$. We note that the effect of screening overwhelms
the effect of NSR and hence the change in the transition temperature.
In particular, with screening included, the NSR effect is even smaller
than with the bare interaction.

The smallness of the NSR effect is consistent with the fact that the
QMC critical temperature $T_c$ of \Refe{Abe2009} satisfies well the
BCS relation $T_c\approx 0.57 \Delta(T=0)$ as can be seen in
\Fig{fig:TstarTc}. Also, the pseudogap computed in \cite{Durel2020} is
very small. Therefore, it is surprising that the temperatures $T^*$ up
to which pair correlations survive in \Refe{Abe2009} can be quite far
above $T_c$. 

%%%%%%%%%%%%%%%%%%%%%%%%%%%%%%%%%%%%%%%%%%%%%%%%%%%%%%%%%%%%%%%%%%%%%%%%%%%%%%%
\section{Triplet pairing}
\label{sect:pwave}
\begin{figure}
 %\begin{center}
  \includegraphics[scale = 0.27, clip = true]{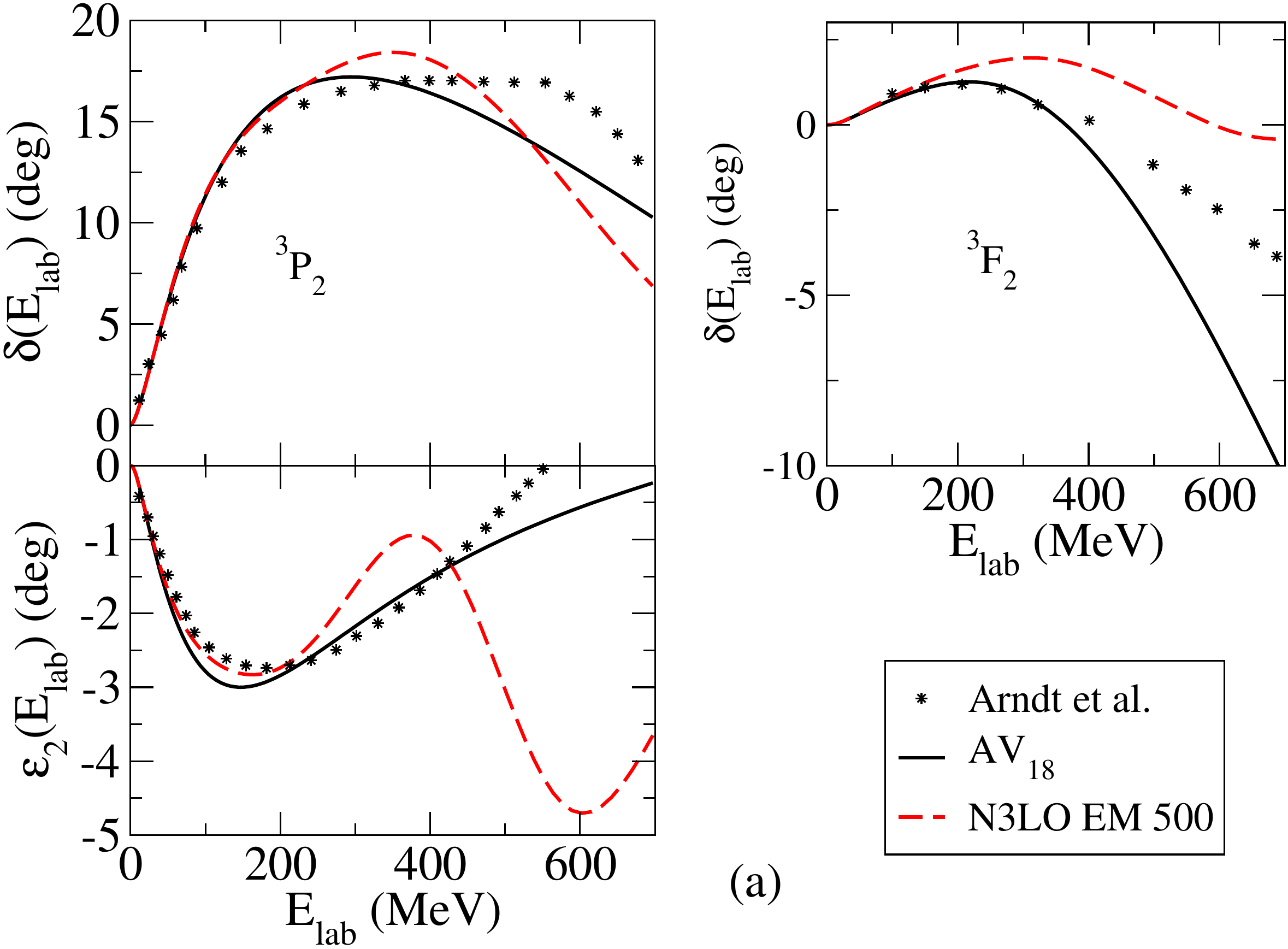} \hspace*{0.1in}
   \includegraphics[scale = 0.25, clip = true]{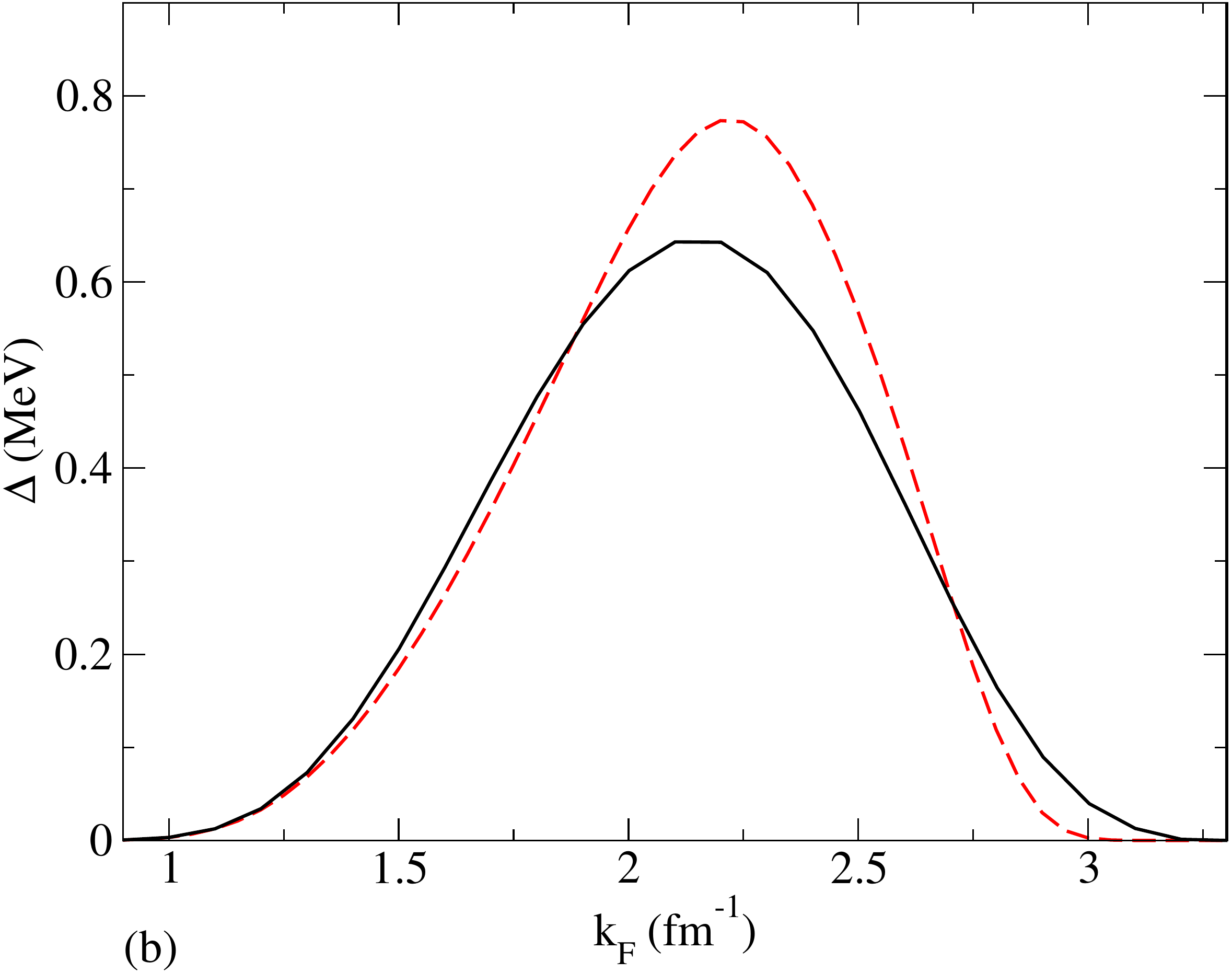}
 %\end{center}
 \caption{Phase shifts and mixing in the $^3PF_2$ channel against the
   experimental phase shift of Arndt et al.~\cite{Arndt1997}. Beyond
   lab energies of $\sim 150 \MeV$, the phase shifts from the
   AV18 and N3LO do not agree with the experimental phase
   shifts. This is reflected in a model dependent gap in at the BCS
   level. It should be noted that the N3LO results for $\kf$ beyond
   $2.5\, \fmi$ becomes unreliable as the chiral cutoff $\Lambda \sim
   3.0\, \fmi$.}
\label{fig:phase-shift-triplet}
\end{figure}

%%%%%%%%%%%%%%%%%%%%%%%%%%%%%%%%%%%%%%%%%%%%%%%%%%%%%%%%%%%%%%%%%%%%%%%%%%%%%%%
Pairing in the triplet channel is supposed to occur at much higher
densities, say, $\kf \gtrsim \, 1.3 \fmi$
(corresponding to number densities $n\gtrsim 0.07 \fm^{-3}$ or mass
densities $\rho \gtrsim 1.2\cdot 10^{14} \text{g}/\text{cm}^3$),
and hence occurs in the outer layers of the neutron star core.
%\footnote{In terms of
%  particle number density the $p$-wave pairing exists in the range
%  $0.07 \, \text{fm}^{-3} \lesssim \rho \lesssim 1.45\,
%  \text{fm}^{-3}$.}.
The evidence for pairing in the spin-triplet channel at high densities
comes from the fact that for momenta $\gtrsim 1.3 \, \fmi$, the
attraction in this channel gets stronger, resulting in
positive two-body phase shifts (see left panel of
Fig.~\ref{fig:phase-shift-triplet}),
%%%%%%%%%%%%%%%%%%%%%%%%%%%%%%%%%%%%%%%%%%%%%%%%%%%%%%%%%%%%%%%%%%%%%%%%%%%%%%%
%%%%%%%%%%%%%%%%%%%%%%%%%%%%%%%%%%%%%%%%%%%%%%%%%%%%%%%%%%%%%%%%%%%%%%%%%%%%%%%%
until it becomes the most attractive channel that supports pairing at
high densities~\cite{Takatsuka1992}. In the spin-triplet channel,
due to the tensor force, the $l = 1$
and $l = 3$ partial waves are coupled, with total angular momentum $J
= 2$, and it is denoted as $^3P_2-^3F_2 \equiv \,^3PF_2$. The zero-temperature
BCS gap is obtained by solving the angle-averaged gap equation~\cite{Baldo1998}
that couples the $l = J \pm 1$ states and is written as,
\begin{equation}
  \Delta_l(k) = -\sum_{l'} \frac{(-1)^{(l - l^\prime)/2}}{\pi}\int_{0}^{\infty}
  q^2dq V_{ll'}(k,q) \frac{\Delta_{l'}(q)}{E(q)},
\label{eq:gap_coup}
\end{equation}
where $E(q) = \sqrt{\xi^2(q) +
  D^2(q)}$ and $\xi(q) = \varepsilon(q) - \mu$. Further, the overlap
between the different partial waves is ignored and $D^2(q) =
\Delta_1^2(q) + \Delta_3^2(q)$~\cite{Takatsuka1992,Baldo1998}. The 
validity of the angle-averaging approximation was confirmed 
in~\cite{Khodel2001}. 

However, pairing in this channel is plagued by uncertainties as the
input free-space two-body interactions, which are the starting point
for the BCS gap equation, are not phase shift
equivalent~\cite{Ding2016,Baldo1998,Maurizio2014,Srinivas2016,Drischler2016,Zuo2008,Papakonstantinou2017}.
This is seen in \Fig{fig:phase-shift-triplet}(a), where the phase
shifts and mixing angle are compared against the experimental phase
shift~\cite{Arndt1997} for two representative
realistic interactions, the phenomenological interaction, AV18~\cite{Wiringa1995} and the chiral
interaction at N3LO~\cite{Entem2003}, as a function of
lab energies. From \Fig{fig:phase-shift-triplet}(a), it is seen that
beyond lab energies of $\approx 150 \, \MeV$, the agreement is rather
poor. These discrepancies result in model dependent gaps already at
the BCS level as seen in \Fig{fig:phase-shift-triplet}(b).

While~\cite{Baldo1998} used realistic interactions to track the model
dependence at the BCS level in the triplet pairing gaps, the input
interactions used in~\cite{Maurizio2014,Srinivas2016,Drischler2016},
are the modern $NN$ interaction obtained via chiral perturation theory
at N3LO which are further softened by the RG running~\cite{Bogner2010}. 
The similarity renormalization group (SRG) interactions ($\vsrg$) are very useful in
studying the gaps in the spin triplet channel. For a given bare
interaction, such as AV18~\cite{Wiringa1995} or
N3LO~\cite{Entem2003}, the SRG evolution preserves the phase shifts at all
energies (unlike $\vlowk$ which preserves the phase shifts only for $k < \Lambda$)
and hence the variation of the gap as a function of the
SRG evolution scale $\lambda$ quantifies the missing $3N$ force and medium corrections,
and, has nothing to do with the inequivalence of the phase
shifts~\cite{Srinivas2016}. In a complementary study, the authors
of~\cite{Maurizio2014}, analysed the dependence of the gap on the
chiral cutoff when the N3LO interactions were used as inputs, which
highlights the differences in dealing with the two pion exchange
interaction term (see Fig.~9 in~\cite{Maurizio2014}).

The correlations beyond the BCS approximation correct both the
quasiparticle spectrum and the particle-particle vertex that enter
the gap equation. The first order Hartree-Fock self-energy 
is given by,
\begin{equation}
  \Sigma^{(1)}(k) = \int \,\frac{d^3k^\prime}{(2\pi)^3} \,
  n_{\vekk^\prime} \sum_{l,S,J} 2 \pi (2J+1) \ip{q}{V_{SllJ}|q} (1
  - (-1)^{l+S+1}),
  \label{eq:self-energy_eqn}
\end{equation}
where $n_{\vekk} = \theta(\kf - k)$ is the Fermi-Dirac
distribution at zero temperature and $q = \vert \vekk - \vekk^\prime
\vert/2$. The HF self-energy changes the free quasiparticle
spectrum to $\varepsilon(k) = k^2/(2 m^*) + \Sigma^{(1)}(k)$, and
$\kf/m^* = [d\varepsilon(k)/dk]_{k = \kf}$ relates the effective mass
to the self-energy.  When $m^* < m$, the density of states near the
Fermi surface decreases and hence one can expect a suppression of
pairing  and therefore, smaller gaps.

\begin{figure}
  \begin{center}
    \includegraphics[scale = 0.3]{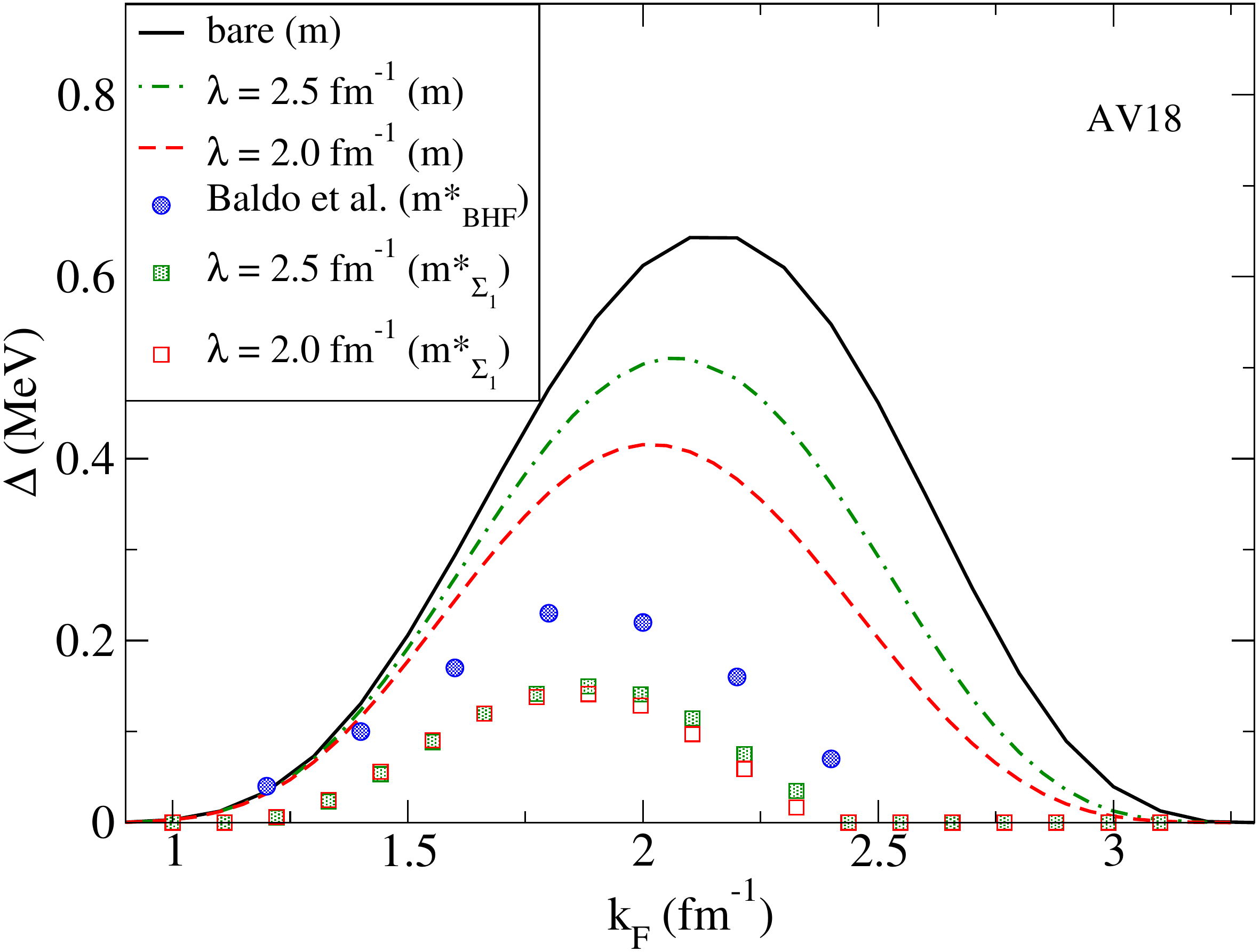}
  \end{center}
  \caption{Medium effects: Comparing the gaps with free single
    particle spectrum (lines) and effective mass (symbols) for the
    AV18 interaction. The blue dots are the results of Baldo et
    al~\cite{Baldo1998} and the squares show the gaps computed with $\vsrg$ for $\lambda
    = 2.5\, \fmi$ (green filled squares) and $\lambda = 2.0 \, \fmi$
    (red empty squares) respectively.}
  \label{fig:effective-mass-3pf2-gaps}
\end{figure}

\Fig{fig:effective-mass-3pf2-gaps}
%%%%%%%%%%%%%%%%%%%%%%%%%%%%%%%%%%%%%%%%%%%%%%%%%%%%%%%%%%%%%%%%%%%%%%%%%%%%%%%
%%%%%%%%%%%%%%%%%%%%%%%%%%%%%%%%%%%%%%%%%%%%%%%%%%%%%%%%%%%%%%%%%%%%%%%%%%%%%%%
shows the gaps with both the free single particle spectrum (lines) and
with the effective mass $m^*$ (symbols - circles and squares). The
black solid line is the bare interaction and the dashed lines and
dash-dotted lines are the gaps obtained from the SRG evolved
interactions for $\lambda = 2.0\, \fmi$ and $2.5\, \fmi$
respectively. The filled circles are the results for the gap with
effective mass calculated from the Br{\"u}ckner Hartree-Fock (BHF) by
Baldo et al.~\cite{Baldo1998}. The squares, filled and empty are the
first order effective mass calculated using
\Eq{eq:self-energy_eqn} for $\lambda = 2.5\, \fmi$ and $2.0\,
\fmi$ respectively. When compared with the free single particle spectrum, the 
inclusion of the effective mass, both BHF and first order, reduces the gaps, due 
to the suppression of the density of states at $\kf$. However, the gaps 
are more suppressed with a first order effective mass compared to the effective mass from BHF. This should be expected at high
densities, as a first order calculation of the self-energy is
insufficient. It is interesting to note that the dependence
on $\lambda$ is dramatically lessened with an effective mass when compared to the corresponding
free spectrum result. Lowering $\lambda$ makes the SRG evolved
interaction more attractive and hence increases the effective
mass. However, including an effective mass reduces the BCS gap. The
dramatic decrease in the $\lambda$ dependence arises due to a
compensation between these two effects.

\begin{figure}
  \begin{center}
    \includegraphics[scale = 0.3,clip = true]{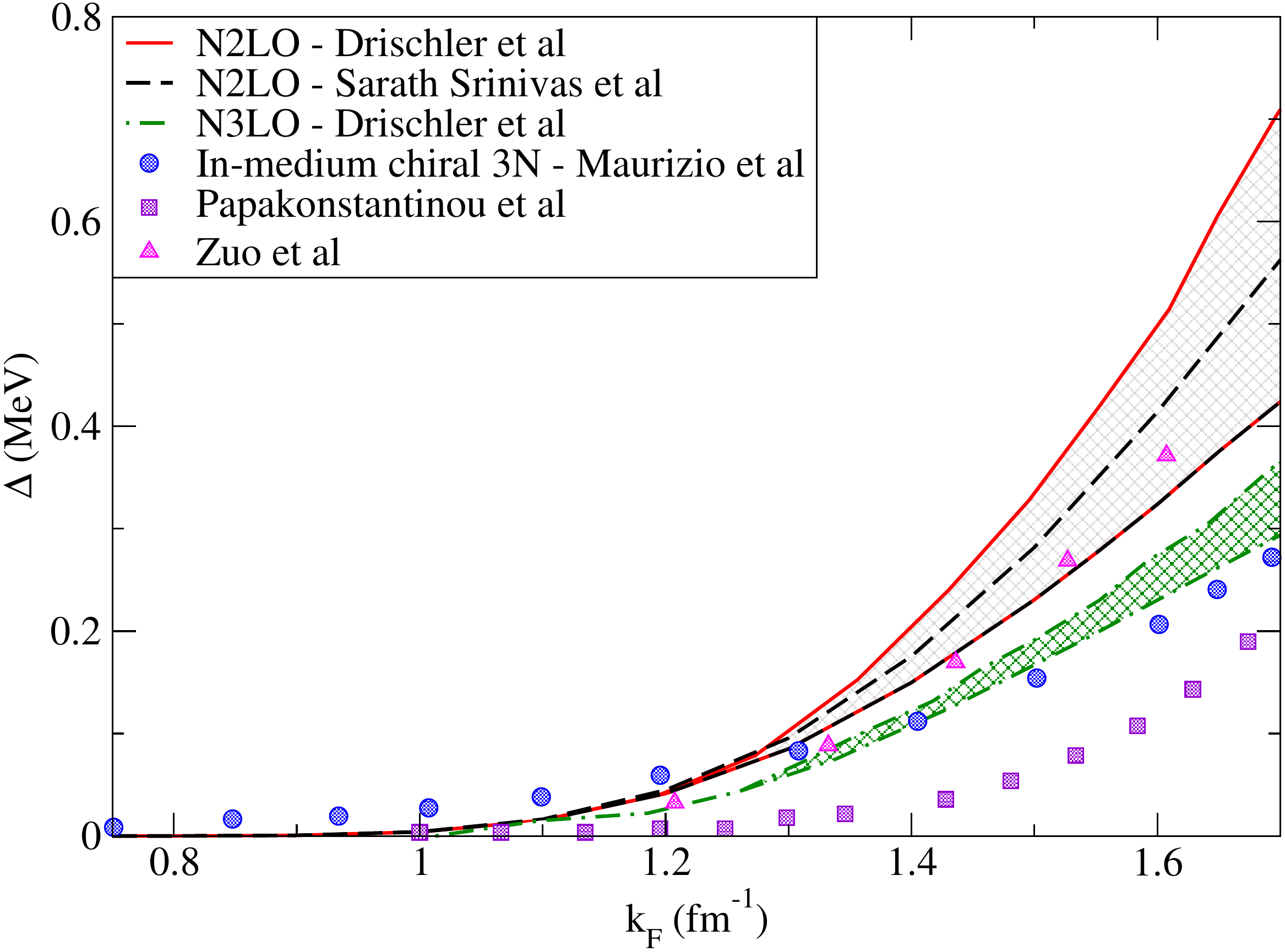}
   \end{center}
  \caption{Triplet gaps including $3N$ interactions}
  \label{fig:triplet-3N} 
\end{figure}

The three-body force is expected to play a crucial role for pairing in the
triplet channel and in fact enhances the
gap~\cite{Maurizio2014,Srinivas2016,Drischler2016,Papakonstantinou2017,Zhou2004,Hebeler2010}.
These forces have been calculated microscopically using
semi-phenomenological
interactions~\cite{Zuo2008,Papakonstantinou2017,Zhou2004,Li2008} as
well as using chiral EFT, where the $3N$ terms first enter at
N2LO~\cite{Hebeler2010,Holt2010}. Fig.~\ref{fig:triplet-3N}
%%%%%%%%%%%%%%%%%%%%%%%%%%%%%%%%%%%%%%%%%%%%%%%%%%%%%%%%%%%%%%%%%%%%%%%%%%%%%%%
%%%%%%%%%%%%%%%%%%%%%%%%%%%%%%%%%%%%%%%%%%%%%%%%%%%%%%%%%%%%%%%%%%%%%%%%%%%%%%%
shows the triplet gap including the $3N$ interaction, which is usually
incorporated as a density dependent $2N$ interaction. While
in~\cite{Maurizio2014}, the density dependent $2N$ interaction is
generated from an in-medium chiral $3N$ force,
\Refs{Srinivas2016,Drischler2016} use an effective density dependent
$2N$ interaction from the $3N$ chiral interactions at N2LO~\cite{Hebeler2010}. In addition,~\cite{Drischler2016} also
considers the $3N$ contributions from N3LO. In \Fig{fig:triplet-3N}, the area shaded in gray
between the solid red lines represents the variations in the gap due
to the uncertainties in the low-energy constants~\cite{Drischler2016}
as well as the three-body cutoff, while the black dashed lines
represent the effect of varying the three-body cutoff after fixing the chiral
low-energy constants~\cite{Srinivas2016}. The green hatched region
between the dash-dotted green lines represents the gaps obtained with $3N$ 
interactions at N3LO with the associated uncertainties in the low-energy
constants~\cite{Drischler2016}. It is worth noting that the spin
triplet gaps are extremely sensitive to the three-body force compared
to the spin singlet
gap~\cite{Drischler2016,Papakonstantinou2017,Hebeler2010}. In the
$^1S_0$ channel the corrections to the gap enter only at higher densities,
while in the $^3PF_2$ channel, the effects of the three-body interaction on
the gap are dramatic.

As discussed already in \Sec{subsect:screening}, beyond BCS correlations (short and long-range) lead to important modifications of the gap. The literature on including these
medium effects for the $p$-wave is rather sparse, with some recent attempts by Dong
et al~\cite{Dong2013} and Ding et al~\cite{Ding2016}. The authors in~\cite{Dong2013}, calculate the quasiparticle weight $Z$ as was done in~\cite{Cao2006} for the singlet
channel (see \Sec{subsect:screening}). The quasiparticle weight
was calculated both with and without the inclusion of a three-body force. The 
presence of the $Z$ factor suppresses the gaps by an entire order of magnitude as well as 
shrinks the density region where the gaps exist.  In~\cite{Ding2016}, the 
short-range correlations are taken into account via the self-consistent Green's function 
techniques, extrapolated to zero temperatures. In addition, the screening corrections as 
in \cite{Cao2006} (see \Sec{subsect:screening}) have been extended to the $p$-wave. In
this case it seems that the screening enhances the gap (antiscreening)
while the short-range correlations suppress it, with the net effect of strongly 
reducing the gaps compared to the BCS results.

While pairing in the triplet channel is an important ingredient to
describe the physics of the neutron star core, much remains to be
explored due to the uncertainties in the free-space interactions.
 
\section{Conclusions}

In this brief review we discussed the state of the art concerning the $s$ and $p$-wave
pairing in pure neutron matter. In the $s$-wave, which is most relevant at low-densities, 
the gap as a function of density seems to be under control. QMC and the most recent many-body calculations agree that the gap, before reaching a maximum of $\sim 2 - 2.5 \MeV$
at $\kf \sim 0.8 \, \fmi$, follows the behavior of the BCS gap reduced by a factor of $0.6$
to $0.7$, except at extremely low-densities (of purely academic interest) where the GMB
(reduction by a factor of $0.45$) limit is reproduced. Effects of BCS-BEC crossover on the
critical temperature seem to be very weak. Beyond the maximum of the gap, 
there are uncertainties that come from medium effects such as effective mass, screening and
short-range correlations.  In addition, there are other factors such as $3N$ 
forces that are important at higher-densities, which we have not discussed here. 

In the $p$-wave, which is supposed to be dominant at high-densities, even at the BCS 
level, the gaps have large uncertainties. Inclusion
of short and long-range correlations seem to reduce the gaps while the $3N$ force 
enhances it. However, since the $p$-wave is always in the extremely weak coupling limit,
the gaps exhibit exponential sensitivity to the details of the interactions and the 
approximations. At densities corresponding to the neutron star core, it would be more 
realistic to consider asymmetric matter with a finite proton fraction. This might 
completely change the conclusions through the $nnp$ $3N$ interactions.  

\vspace*{0.1in} 
\noindent \textbf{Acknowledgment:} The authors acknowledge support from Collaborative Research Program of IFCPAR/CEFIPRA, Project number: 6304-4. 

\vspace*{0.1in} 
\noindent \textbf{Author contribution statement:} Both authors contributed equally to this work.

%%%%%%%%%%%%%%%%%%%%%%%%%%%%%%%%%%%%%%%%%%%%%%%%%%%%%%%%%%%%%%%%


\begin{thebibliography}{99}
\bibitem{Migdal1959}
  A.~B.~Migdal,
  Zh. Eksp. Teor. Fiz. {\bf 37} 249 (1959)
  [Sov. Phys. JETP {\bf 10}, 176 (1960)]; Nucl. Phys. {\bf 13}, 655 (1959).
  %``Superfluidity and the moments of inertia of nuclei''
\bibitem{Ginzburg1964}
  V.~L.~Ginzburg and D.~A.~Kirzhnits,
  Zh. Eksp. Teor. Fiz. {\bf 47}, 2006 (1964)
  [Sov. Phys.  JETP {\bf 20}, 1346 (1965)].
  %``On the superfluidity of neutron stars''
\bibitem{Bell1968}
  J. Bell, A. Hewish, J. D. H. Pilkington, P. F. Scott, and R. A. Collins,
  Nature \textbf{217}, 709 (1968).
  %``Observation of a Rapidly Pulsating Radio Source.''
\bibitem{Gold1968}
  T. Gold,
  Nature {\bf 218}, 731 (1968). 
  %``Rotating neutron stars as the origin of the pulsating radio sources.''
\bibitem{Baym1969}
  G. Baym, C. J. Pethick, D. Pines, and Malvin Ruderman,
  Nature {\bf 224}, 872 (1969).
  %``Spin Up in Neutron Stars : The Future of the Vela Pulsar''
\bibitem{Pines1985}
  D. Pines and M. A. Alpar,
  Nature {\bf 316}, 27 (1985).
  %``Superfluidity in neutron stars''
\bibitem{Andersson2012} N. Andersson, K. Glampedakis, W. C. G. Ho, and
  C. M. Espinoza Phys. Rev. Lett. {\bf 109}, 241103 (2012).
  %``Pulsar Glitches: The Crust is not Enough''
\bibitem{Yakovlev2004}
  D.~G.~Yakovlev and C.~J.~Pethick,
  Ann.\ Rev.\ Astron.\ Astrophys.\ {\bf 42}, 169 (2004).
  %``Neutron-star cooling''
  %doi:10.1146/annurev.astro.42.053102.134013 [astro-ph/0402143].
\bibitem{Page2009}
  D.~Page, J.~M.~Lattimer, M.~Prakash and A.~W.~Steiner,
  Astrophys.\ J.\  {\bf 707}, 1131 (2009).  
  %``Neutrino Emission from Cooper Pairs and Minimal Cooling of Neutron Stars,''
\bibitem{Anderson1975}
  P.~W.~Anderson and N.~Itoh,
  Nature {\bf 256}, 25 (1975).
  %``Pulsar glitches and restlessness as a hard superfluidity phenomenon''
\bibitem{Bohr1958}
  A. Bohr, B.~R.~Mottelson and D.~Pines,
  Phys.~Rev.\textbf{110}, 936, (1958).
\bibitem{BohrMottelson1} A.~Bohr and B.~R.~Mottelson,
  \textit{Nuclear Structure}, Vol. 1 (Benjamin, New York, 1959).
\bibitem{Fang2004}
  D. Q. Fang \textit{et al.},
  Phys. Rev. C \textbf{69}, 034613, (2004).
\bibitem{Ozawa2001}
  A. Ozawa \textit{et al.},
  Nucl. Phys. A \textbf{691}, 599, (2001).
\bibitem{Hagino2011}
  K.~Hagino and H.~Sagawa,
  Phys. Rev. C \textbf{84}, 011303 (2011).
\bibitem{Chamel2008}
  N. Chamel and P. Haensel,
  Living Rev. Relativity, 11, 10 (2008).
\bibitem{Dean2003}
  D.~J.~Dean and M.~Hjorth-Jensen,
  Rev.\ Mod.\ Phys.\ {\bf 75}, 607 (2003).  
\bibitem{Haskell2018}
  B.~Haskell and A.~Sedrakian, in: L.~Rezzolla, P. Pizzochero, D.~Jones, N.~Rea, I. Vida\~na (eds.) {\it The Physics and Astrophysics of Neutron Stars.} Astrophysics and Space Science Library, Vol. 457 (Springer, Cham, 2018).
%  %``Superfluidity and Superconductivity in Neutron Stars,''
\bibitem{Sedrakian2019} 
  A.~Sedrakian and J.~W.~Clark,
  Eur.\ Phys.\ J.\ A {\bf 55}, 167 (2019).
  %``Superfluidity in nuclear systems and neutron stars,''
  %doi:10.1140/epja/i2019-12863-6
  %[arXiv:1802.00017 [nucl-th]].
\bibitem{Schrieffer}
J. R. Schrieffer, Theory of Superconductivity (Benjamin, New York, 1964).  
\bibitem{Bogner2007}
  S. K. Bogner, R. J. Furnstahl, S. Ramanan, and A. Schwenk,
  Nucl. Phys. A \textbf{784}, 79 (2007).
\bibitem{Bogner2010} S. Bogner, R. Furnstahl, and A. Schwenk,
  Prog. Part. Nucl. Phys. \textbf{65}, 94 (2010).
  %``From low-momentum interactions to nuclear structure.''
\bibitem{Hebeler2007}
  K. Hebeler, A. Schwenk, and B. Friman,
  Phys. Lett. B \textbf{648} 176 (2007).
  % ``Dependence on the 1S0 superfluid pairing gap on nuclear interactions''
  %DOI: 10.1016/j.physletb.2007.03.022
\bibitem{Buraczynski2019}
    M.~Buraczynski, N.~Ismail, and A.~Gezerlis,
  Phys. Rev. Lett. {\bf 122}, 152701 (2019).
  %``Nonperturbative Extraction of the Effective Mass in Neutron Matter''
\bibitem{Decharge1980}
  J. Decharg\'e and D. Gogny,
  Phys. Rev. C {\bf 21}, 1568 (1980).
\bibitem{Chappert2008}
  F. Chappert, M. Girod, and S. Hilaire,
  Phys. Lett. B \textbf{668}, 420 (2008).
\bibitem{Chabanat1998}
  E. Chabanat, P. Bonche, P. Haensel, J. Meyer, and R. Schaeffer,
  Nucl. Phys. A \textbf{635}, 231 (1998).
\bibitem{Chamel2009}
  N. Chamel, S. Goriely, and J.M. Pearson,
  Phys. Rev. C \textbf{80}, 065804 (2009).
\bibitem{Goriely2010}
  S. Goriely, N. Chamel, and J.M. Pearson,
  Phys. Rev. C \textbf{82}, 035804 (2010).
\bibitem{Gorkov1961}
  L. P. Gor'kov and T.K. Melik-Barkhudarov,
  Zh. Eksp. Teor. Fiz. \textbf{40}, 1452 (1961)
  [Sov. Phys. JETP \textbf{13}, 1018 (1961)].
\bibitem{Wambach1993}
  J.~Wambach, T.~L.~Ainsworth and D.~Pines,
  Nucl.\ Phys.\ A {\bf 555}, 128 (1993).
  %doi:10.1016/0375-9474(93)90317-Q
\bibitem{Schulze1996}
  H.~J.~Schulze, J.~Cugnon, A.~Lejeune, M.~Baldo and U.~Lombardo,
  Phys.\ Lett.\ B {\bf 375}, 1 (1996).
  %doi:10.1016/0370-2693(96)00213-4
\bibitem{Shen2003}
  C.~Shen, U.~Lombardo and P.~Schuck,
  Phys.\ Rev.\ C {\bf 67}, 061302 (2003).
  %doi:10.1103/PhysRevC.67.061302 [nucl-th/0212027].
\bibitem{Shen2005}
  Caiwan Shen, U. Lombardo, and P. Schuck,
  Phys. Rev. C \textbf{71}, 054301 (2005).
\bibitem{Cao2006}
  L. G. Cao, U. Lombardo, and P. Schuck,
  Phys. Rev. C \textbf{74}, 064301 (2006).
\bibitem{Ramanan2018}
  S. Ramanan and M. Urban, Phys. Rev. C
  \textbf{98}, 024314 (2018).
  %``Screening and antiscreening of the pairing interaction in low-density neutron matter''
\bibitem{Urban2020}
  M.~Urban and S.~Ramanan,
  Phys. Rev. C \textbf{101}, 035803 (2020).
  %``Neutron pairing with medium polarization beyond the Landau approximation,''
\bibitem{Gandolfi2008}
  S. Gandolfi, A. Yu. Illarionov, S. Fantoni, F. Pederiva, and K. E. Schmidt,
  Phys. Rev. Lett. \textbf{101}, 132501, (2008).
\bibitem{Abe2009}
  T.~Abe and R.~Seki,
  Phys. Rev. C \textbf{79}, 054002 (2009).
\bibitem{Gezerlis2010}
  A. Gezerlis and J. Carlson, Phys. Rev. C
  \textbf{81}, 025803 (2010).
\bibitem{FetterWalecka}
  A. L. Fetter and J. D. Walecka,
  \textit{Quantum Theory of Many-Particle Systems}
  (McGraw-Hill, New York, 1971).
\bibitem{Baldo2000}
	M.~Baldo and A.~Grasso, Phys. Lett. B \textbf{485}, 115 (2000).  
\bibitem{Babu1973}
  S.~V.~Babu and G.~E.~Brown,
  Ann. Phys. (N.Y.) \textbf{77}, 1 (1973).
\bibitem{Garcia1992}
  C. Garc\'ia-Recio, J. Navarro, Van Giai Nguyen, and L.L. Salcedo,
  Ann. Phys. (N.Y.) \textbf{214}, 293 (1992).
\bibitem{Pastore2015}
  A. Pastore, D. Davesne, and J. Navarro,
  Phys. Rep. \textbf{563}, 1 (2015).
\bibitem{Ding2016}
  D. Ding, A. Rios, H. Dussan, W. H. Dickhoff, S. J. Witte, A. Carbone,
  and A. Polls,
  Phys. Rev. C \textbf{94}, 025802 (2016).
\bibitem{Pavlou2017}
G.~E.~Pavlou, E.~Mavrommatis, C.~Moustakidis and J.~W.~Clark,
%``Microscopic Study of ${}^1{S_0}$ Superfluidity in Dilute Neutron Matter,''
Eur. Phys. J. A \textbf{53}, 96 (2017).  
\bibitem{CalvaneseStrinati2018}
  G.~Calvanese Strinati, P.~Pieri, G.~R\"opke, P.~Schuck, and M.~Urban,
  Phys. Rep. {\bf 738}, 1 (2018).
  %``The BCS-BEC crossover: From ultra-cold Fermi gases to nuclear systems''
\bibitem{Ohashi2020} Y. Ohashi, H. Tajima, and P. van Wyk,
  %BCS-BEC crossover in cold atomic and in nuclear systems
  Prog. Part. Nucl. Phys. \textbf{111}, 103739 (2020).
\bibitem{Baldo1995}
  M. Baldo, U. Lombardo, P. Schuck,
  Phys. Rev. C \textbf{52}, 975 (1995).
  %Deuteron formation in expanding nuclear matter from a strong coupling BCS approach,
\bibitem{Matsuo2006}
  M. Matsuo,
  Phys. Rev. C \textbf{73}, 044309 (2006).
\bibitem{Margueron2007}
  J. Margueron, H. Sagawa, and K. Hagino,
  Phys. Rev. C \textbf{76}, 064316 (2007).
\bibitem{Nozieres1985}
  P. Nozi{\`e}res and S. Schmitt-Rink,
  J. Low. Temp. Phys. \textbf{59}, 195 (1985).
\bibitem{Zimmermann1985}
  R. Zimmermann and H. Stolz,
  Phys. Status Solidi B \textbf{131}, 151 (1985).
\bibitem{Schmidt1990}
  M. Schmidt, G. R{\"o}pke, H. Schulz,
  Ann. Phys. \textbf{202}, 57 (1990).
  %``Generalized Beth-Uhlenbeck approach for hot nuclear matter,''
\bibitem{Jin2010}
  M.~Jin, M.~Urban and P.~Schuck,
  Phys. Rev. C \textbf{82}, 024911 (2010).    
  %``BEC-BCS Crossover and the Liquid-Gas Phase Transition in Hot and Dense Nuclear Matter,''
\bibitem{Ramanan2013}
  S.~Ramanan and M.~Urban,
  Phys.\ Rev.\ C {\bf 88}, 054315 (2013).
\bibitem{Tajima2019}
  H. Tajima, T. Hatsuda, P. van Wyk, and Y. Ohashi,
  Scientific Reports \textbf{9}, 18477 (2019).
  %Superfluid Phase Transition and Effects of Thermal Pairing Fluctuations in Asymmetric Nuclear Matter,
\bibitem{Durel2020}
  D. Durel and M. Urban,
  preprint arxiv:2010.04672 [nucl-th] (2020).
\bibitem{Arndt1997}
  R.~A.~Arndt, C.~H.~Oh, I.~I.~Strakovsky, R.~L.~Workman, and F.~Dohrmann,
  Phys. Rev. C \textbf{56}, 3005 (1997).
  %experimental phase shifts
\bibitem{Takatsuka1992}
  T.~Takatsuka and R.~Tamagaki,
  Prog. Theor. Phys. Suppl. \textbf{112}, 27-66 (1993).
  %``Superfluidity in neutron star matter and symmetric nuclear matter,''
\bibitem{Baldo1998}
  M. Baldo, {\O}. Elgar{\o}y, L. Engvik, M. Hjorth-Jensen, and H.-J. Schulze, Phys. Rev. C \textbf{58}, 1921 (1998).
  %``Triplet P-3(2) to F-3(2) pairing in neutron matter with modern nucleon-nucleon potentials,''
\bibitem{Khodel2001}
V. V. Khodel, V. A. Khodel, and J. W. Clark, Nucl. Phys. A \textbf{679}, 827 (2001).
\bibitem{Maurizio2014}
  S.~Maurizio, J.~W.~Holt and P.~Finelli,
  Phys.\ Rev.\ C {\bf 90} 044003 (2014).
  %``Nuclear pairing from microscopic forces: singlet channels and higher-partial waves,''
\bibitem{Srinivas2016} S.~Srinivas and S.~Ramanan,
  Phys.\ Rev.\ C {\bf 94}, 064303 (2016).
  %``Triplet Pairing in pure neutron matter,''
\bibitem{Drischler2016}
  C.~Drischler, T.~Kr{\"u}ger, K.~Hebeler and A.~Schwenk,
  Phys.\ Rev.\ C {\bf 95}, 024302 (2017).  
\bibitem{Zuo2008}
  W.~Zuo, C.~X.~Cui, U.~Lombardo and H.~J.~Schulze,
  Phys. Rev. C \textbf{78} (2008).
  %``Three-body force effect on P-3 F-2 neutron superfluidity in neutron matter, neutron star matter, and neutron stars,''
\bibitem{Papakonstantinou2017}
  P.~Papakonstantinou and J.~W.~Clark,
  J.\ Low.\ Temp.\ Phys.\ {\bf 189}, 361 (2017).
\bibitem{Wiringa1995}
  R.~B.~Wiringa, V.~G.~J.~Stoks, and R.~Schiavilla
  Phys.\ Rev.\ C {\bf 51}, 38–51 (1995). 
  %Accurate nucleon-nucleon potential with charge-independence breaking
\bibitem{Entem2003}
  D.~R.~Entem and R.~Machleidt,
  Phys.\ Rev.\ C {\bf 68}, 041001 (2003).
  %``Accurate charge dependent nucleon nucleon potential at fourth order of chiral perturbation theory,''
  %%CITATION = NUCL-TH/0304018;%%
\bibitem{Zhou2004}
  X.~R.~Zhou, H.~J.~Schulze, E.~G.~Zhao, F.~Pan and J.~P.~Draayer,
  Phys. Rev. C \textbf{70}, 048802 (2004).
  %``Pairing gaps in neutron stars,''
\bibitem{Hebeler2010}
  K.~Hebeler and A.~Schwenk,
  Phys. Rev. C \textbf{82}, 014314 (2010).
  %``Chiral three-nucleon forces and neutron matter,''
  %doi:10.1103/PhysRevC.82.014314
\bibitem{Li2008}
  Z. H. Li, U. Lombardo, H.-J. Schulze, and W. Zuo,
  Phys.~Rev.~C \textbf{77}, 034316 (2008).
  %3N interaction
\bibitem{Holt2010}
  J.~W.~Holt, N.~Kaiser and W.~Weise, 
  Phys. Rev. C \textbf{81}, 024002 (2010).
  %Density-dependent effective nucleon-nucleon interaction from chiral three-nucleon forces,
\bibitem{Dong2013}
  J.~M.~Dong, U.~Lombardo and W.~Zuo,
  Phys. Rev. C \textbf{87}, 062801 (2013).
  %``$^3PF_2$ pairing in high-density neutron matter,''


\end{thebibliography}
\end{document}